\documentclass[dvips]{acta}
\usepackage{supertabular,lscape,epsfig}
\usepackage{amssymb}
\usepackage{amsmath}
\usepackage{color}
\usepackage{graphicx}
\usepackage{latexsym}

\SetPages{0}{0}

\SetVol{0}{2018}

\newcommand{\teff}{$T_{\rm eff}$}
\newcommand{\logg}{$\log g$}
\newcommand{\vsini}{$v\sin i$}

\newcommand{\kms}{km\,s$^{-1}$}
\newcommand{\fluxunit}{erg\,cm$^{-2}$s$^{-1}$\AA$^{-1}$} 

\newcommand{\Msun}{{$M_{\odot}$}}
\newcommand{\Lsun}{{$L_{\odot}$}}
\newcommand{\Rsun}{{$R_{\odot}$}}

\newcommand{\Mstar}{{$M_{\star}$}}

\newcommand{\Ll}{{$L_{\rm line}$}}

\newcommand{\Lacc}{{$L_{\rm acc}$}}
\newcommand{\Macc}{{$\dot{M}_{\rm acc}$}}

\begin{document}

\begin{Titlepage}
\Title{Contemporaneous broad-band photometry and H$\alpha$ observations of T~Tauri stars\footnote{Based on observations 
made with the 0.91-m telescope of Catania Astrophysical Observatory (Italy) and with the INT telescope operated on the island of La Palma by 
the Isaac Newton Group of Telescopes in the Spanish Observatorio del Roque de los Muchachos of the Instituto 
de Astrofisica de Canarias.}
}
\Author{A.~F~r~a~s~c~a$^1$, D. M~o~n~t~e~s$^2$, J.M.~A~l~c~a~l~\'a$^3$, A.~K~l~u~t~s~c~h$^1$ and 
P.~G~u~i~l~l~o~u~t$^4$}
{$^1$ INAF - Osservatorio Astrofisico di Catania, via S. Sofia, 78, I-95123 Catania, Italy\\
e-mail: antonio.frasca@oact.inaf.it\\
$^2$ Departamento de F\'{\i}sica de la Tierra y Astrof\'{\i}sica \& UPARCOS-UCM (Unidad de F\'{\i}sica de Part{\'\i}culas y del Cosmos de la UCM), Facultad de Ciencias F\'{\i}sicas, Universidad Complutense de Madrid, E-28040 Madrid, Spain\\
e-mail: dmontes@ucm.es\\
$^3$ INAF-Osservatorio Astronomico di Capodimonte, via Moiariello 16, I-80131 Napoli, Italy\\
e-mail: jmae@na.astro.it\\
$^4$ Universit\'e de Strasbourg, CNRS, Observatoire astronomique de Strasbourg, UMR 7550, F-67000 Strasbourg, France\\
e-mail: patrick.guillout@astro.unistra.fr}

\Received{Month Day, Year}
\end{Titlepage}

\Abstract{The study of contemporaneous variations of the continuum flux and emission lines is of great importance
to understand the different astrophysical processes at work in T~Tauri stars.

In this paper we present the results of a simultaneous $BVRI$ and H$\alpha$ photometric monitoring, contemporaneous
to medium-resolution spectroscopy of six T~Tauri stars in the Taurus-Auriga star forming region. 
We have characterized the H$\alpha$ photometric system using synthetic templates and the contemporaneous 
spectra of the targets. We show that we can achieve a precision corresponding to  2--3\,\AA\ in the H$\alpha$ equivalent 
width, in typical observing conditions.  

The spectral analysis has allowed us to determine the basic stellar parameters and the 
values of quantities related to the accretion. In particular, we have measured a significant veiling only for the
three targets with the strongest H$\alpha$ emission (T~Tau, FM~Tau, and DG~Tau).

The broad-band photometric variations are found to be in the range 0.05--0.70 mag and are often paired to variations in
the H$\alpha$ intensity, which becomes stronger when the stellar continuum is weaker. 
In addition, we have mostly observed a redder $V-I$ and a bluer $B-V$ color as the stars become fainter.
For most of the targets, the timescales of these variations seem to be longer than the rotation period. 
One exception is T~Tau, for which the broad-band photometry varies with the rotation period.
The most plausible interpretation of these photometric and H$\alpha$ variations is that they are due to non-stationary 
mass accretion onto the stars, but rotational modulation can play a major role in some cases.}{Stars: pre-main sequence, 
activity -- Accretion, accretion disks -- Open clusters and associations: Taurus}

\section{Introduction}
\label{Sec:intro}

Young pre-main sequence (PMS) stars display a rich variety of phenomena in their atmospheres and circumstellar 
environments, including magnetic activity, mass accretion from circumstellar disks, outflows, and jets, which produce 
typical signatures in their spectra and energy distributions. 
However, different phenomena can produce similar effects on the observed properties of PMS stars or can affect 
the same spectral features. Indeed, most diagnostics of mass accretion from the disk onto the 
central object for the classical T~Tauri stars (CTTs) are emission lines that are also diagnostics of chromospheric 
activity for weak-lined T~Tauri stars (WTTs). The latter, however, display both much weaker emission lines 
and significantly lower or no infrared excess than the former, hence undergo much less or no accretion. 
The boundary between weak accretors and WTTs is not sharp, as chromospheric emission is always present in PMS stars,
and may provide a non-negligible contribution to the excess emission with respect to the photospheric flux in weak 
accretors (see, e.g., Manara et al. 2013, Ingleby et al. 2011, Frasca et al. 2015, Alcal\'a et al. 2017, and references therein).

An important feature of PMS stars is the variability of both their continuum flux and emission lines with both short
and long timescales (e.g., Cody et al. 2014, Costigan et al. 2014, and references therein).
The photometric variations in different bands carry important information on the 
physical mechanisms operating in the PMS stars and in their circumstellar environments.
Rotation periods can be determined for both the CTTs and the WTTs.
Indeed, due to the rotation of the star, the stellar flux can be modulated by hot spots that are produced
by the impact of the accretion flow on the stellar photosphere, for the strongly accreting CTTs, or to photospheric dark spots 
related to the stellar magnetic fields for both classes of objects.
In addition, flux variations on time scales different than the star rotation period can arise from variable accretion and outflows, 
and obscuration by the circumstellar disk (e.g., Bouvier et al. 2007, Rodriguez et al. 2017).  

High-cadence, long-term photometry, like that obtained using space telescopes, is able to reveal different types of 
variability with various time scales (periodic variations, dips, bursts, stochastic variations, etc.), which are interpreted as the result of different 
phyisical processes and geometric effects  (e.g., Cody et al. 2014, Sousa et al. 2016, Venuti et al. 2017). 
Multiband photometry in optical and near-infrared (NIR) bands is useful, for the periodic variables, to distinguish between hot and 
cool spots as the source of the modulation, and to determine their mean properties (e.g., Bouvier et al. 1993, Frasca et al. 2009).  

Despite the ever growing number of photometric data of high quality as regards precision and cadence, 
there are few works in the literature based on multiband photometric observations taken contemporaneously
to spectroscopy (e.g., Bouvier et al. 2003, Alencar et al. 2012, Sousa et al. 2016). 
The study of contemporaneous variations of the continuum flux and emission spectral lines, which are  
diagnostics of jets, accretion, and/or chromospheric activity, is of great importance to get a better understanding of the 
phenomena at work in T~Tauri stars. Moreover, the acquisition of highly-accurate flux-calibrated spectra is very
difficult, hence contemporaneous photometric observations are crucial to overcome this problem, allowing the conversion of 
emission-line equivalent widths into line fluxes.
Furthermore, narrow-band photometry is a very efficient tool to obtain information on line fluxes even in the case of
very faint objects, in which spectroscopic observations are normally not possible or would provide very low signal-to-noise useless 
data. 

In this paper we present a study of six T~Tauri stars in the Taurus-Auriga complex, which is based on 
simultaneous $BVRI$ and narrow-band H$\alpha$ photometry. The photometry, gathered with a 0.9-m telescope, has been  
contemporaneously acquired to intermediate-resolution spectra that include the H$\alpha$ line. We present and discuss the 
characteristics and performances of the H$\alpha$ photometric system that can be used for larger observational campaigns.

The spectra are used not only to check and calibrate the H$\alpha$ photometry, but also to derive physical and accretion 
parameters for these sources.

%
\section{Observations and data reduction}
\label{Sec:Observations}
We selected a small sample of six bright young stars in the nearby star forming region Taurus-Auriga 
($d\simeq 140$\,pc, Kenyon et al. 1994). Table~1 lists the relevant information reported in the literature 
regarding their spectral type (SpType), which ranges from G2 to M0, their maximum $V$ magnitude (in the range 9--12 mag), and 
other stellar parameters.

\subsection{Photometry}
\label{Sec:Photometry}
The photometric observations have been carried out at the M. G. Fracastoro station (Serra La Nave, Mt. Etna,
1750 m a.s.l.) of the Osservatorio Astrofisico di Catania (OAC -- Italy) from 19 to 24 November 2014. 
We used the the imaging camera with the 91-cm telescope and a set of broad-band filters (Johnson-Cousins 
$B$, $V$, $R$, $I$) as well as two narrow band filters.	
The CCD camera adopts a Kodak KAF1001E 1k$\times$1k chip that, with the focal reducer, covers a field of view of about 
13$\times$13 arcminutes.\footnote{\scriptsize \tt http://sln.oact.inaf.it/sln\_old/dmdocuments/ccd91rappint2-07.pdf}

\begin{figure}[htb]
\includegraphics[width=9cm]{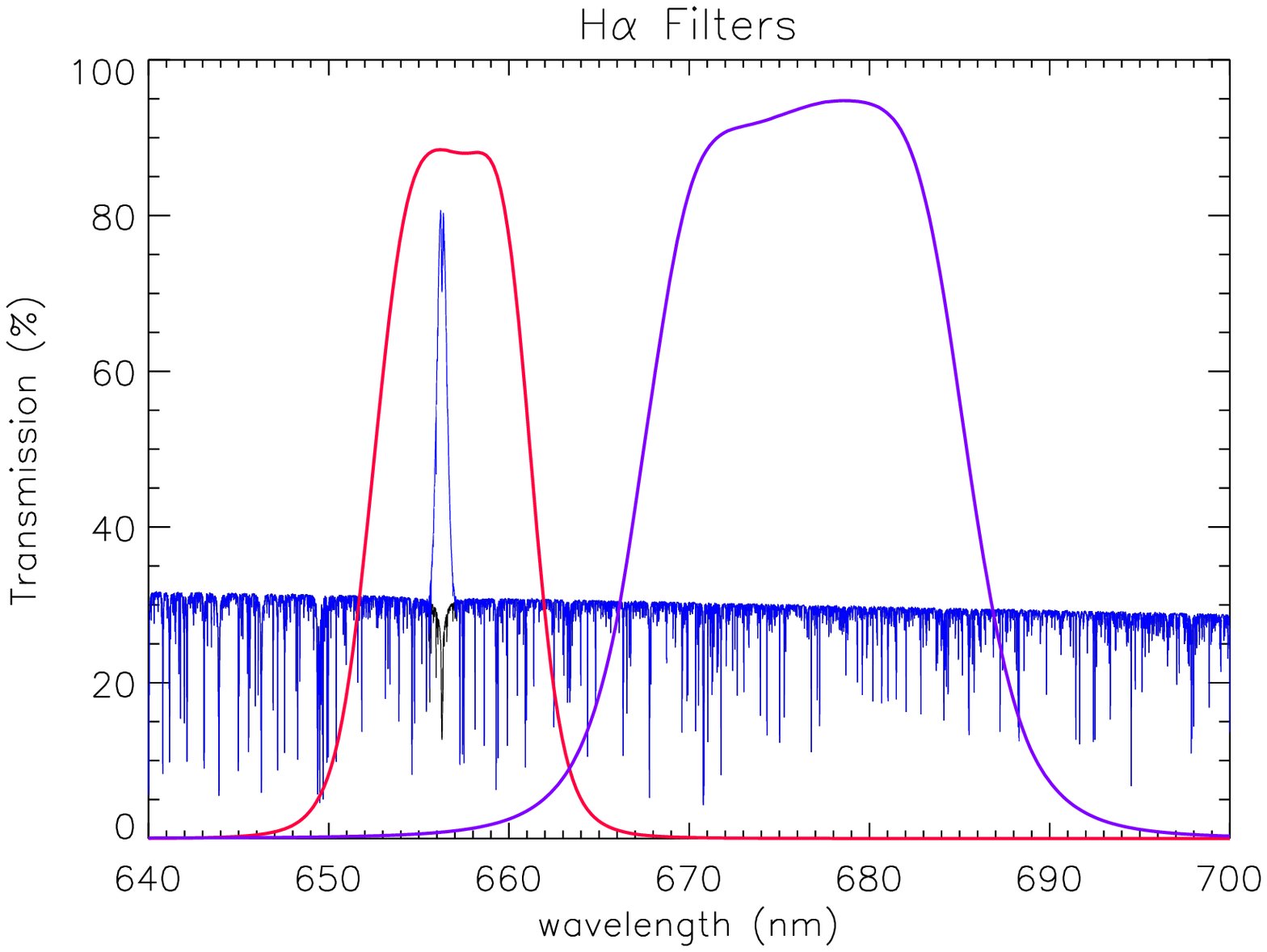}
\FigCap{Passbands of the H$\alpha$ filters (red for $H\alpha_9$ and purple for $H\alpha_{18}$, respectively) superimposed to 
a synthetic BT-Settl spectrum with \teff=5000\,K and \logg=4.0 (black line). We also overplot a simulated CTT 
spectrum (blue line) that is obtained adding a Gaussian with FWHM=6\,\AA\ and an area EW=12\,\AA\ to the same synthetic spectrum.}
\end{figure}

The two narrow-band filters, $H\alpha_9$ and $H\alpha_{18}$, were manufactured by ASAHI SPECTRA\footnote{\tt http://www.asahi-spectra.com/}. 
They have central wavelengths of 656.8\,nm and  676.4\,nm and full-width at half maximum (FWHM) of 9 and 18\,nm, respectively.
The first filter is centered at the H$\alpha$ wavelength, while the second one collects radiation from a continuum region
adjacent to the H$\alpha$. The passbands of these two filters are shown in Fig.~1 overplotted to
a BT-Settl synthetic spectrum (Allard et al. 2012) with a solar iron abundance, \teff=5000\,K, and \logg=4.0. The same figure shows a 
template spectrum that is obtained by summing a Gaussian H$\alpha$ emission profile with FWHM=6\,\AA\ (274\,\kms) and equivalent width 
$EW_{\rm H\alpha}$\,=\,12\,\AA\  to the same BT-Settl spectrum. 

\MakeTable{llrllccc}{12.5cm}{Properties of the observed T~Tauri stars.}
{\hline
\noalign{\smallskip}
Name       	  & SpType   & $V_{\rm max}$   & Period  & Dist  & Mass$^{a}$ & Radius$^{a}$ \\  
           	  &	     &  (mag)          & (day)   & (pc)  &  (\Msun)   & (\Rsun)      \\ 		     
\noalign{\smallskip}
\hline
\noalign{\smallskip}
T~Tau       & K0\,IV--Ve  &  9.8$^{b}$  &  2.81$^{f}$ & 139.1$^{i}$  &  1.99  &  3.73  \\      
RY~Tau     & K1\,IV--Ve  & 10.0$^{c}$  &  5.6 $^{f}$ & 176.7$^{i}$  &  2.11  &  3.10  \\   
V773~Tau & K3\,Ve  	& 10.7$^{b}$  &  3.08$^{g}$ & 138.1$^{i}$  &  0.98  &  3.03  \\  	
FM~Tau    & M0--2e	& 12.0$^{d}$  &  3.07$^{h}$ & 140  $^{j}$  &  0.15  &  0.97  \\     
DG~Tau    & K6\,Ve	& 11.8$^{e}$  &  6.3 $^{f}$ & 140  $^{j}$  &  0.75  &  1.58  \\   
SU~Aur    & G2IIIne 	&  8.9$^{c}$  &  2.66$^{f}$ & 142.4$^{i}$  &  2.07  &  2.72  \\  
\hline
\noalign{\smallskip}
\multicolumn{8}{p{9cm}}{$^{a}$ = Herczeg \& Hillenbrand 2014; $^{b}$ = ESA 1997$^{c}$ = Herbig et al. 1988.}\\ 
\multicolumn{8}{p{9cm}}{ $^{d}$ = Alfonso-Garzon et al. 2012; $^{e}$ = Hauck et al. 1990.}\\
\multicolumn{8}{p{9cm}}{$^{f}$ = Watson et al. 2015; $^{g}$ = Norton et al. 2007; $^{h}$ = Davies et al. 2014.}\\
\multicolumn{8}{p{9cm}}{$^{i}$ = Gaia Collaboration 2016; $^{j}$ = Kenyon et al. 1994.}\\
}

The exposure times ranged from 2 to 60 s  depending on the filter and the star brightness. For each target, we typically acquired 
one or two sequences of $BVRI$ exposures per night followed by a sequence of three exposures $H\alpha_{18}$--$H\alpha_9$--$H\alpha_{18}$.  

Data reduction was carried out following standard steps of overscan region subtraction, master-bias subtraction, 
and division by average twilight flat-field images. 
The magnitudes were measured on the corrected images through aperture photometry performed with ad-hoc
IDL procedures based on DAOPHOT. We extracted the instrumental magnitudes using apertures with a different radius.
To minimize the effects of seeing variations from one night to another (in the range 1.5"--3.0") we have chosen the aperture
with a radius of 8 pixels (6 arcsec).  
The photometric errors related to the photon statistics are in the range 0.001--0.005\,mag in all bands for all the targets with the exception of FM~Tau,
for which they are of about 0.01\,mag. Indeed, this target is included in the field of the brighter component V773~Tau that required
shorter exposures to avoid saturation.
A more reliable evaluation of the precision of our measurements is provided by the standard deviation of the differential magnitudes between
two non-variable stars in the target's field. This value is in the range 0.010--0.025\,mag for stars with a similar, or slightly
fainter, brightness as the targets (see Fig.~2).	

\begin{figure}[htb]
\includegraphics[width=6.8cm,height=6cm]{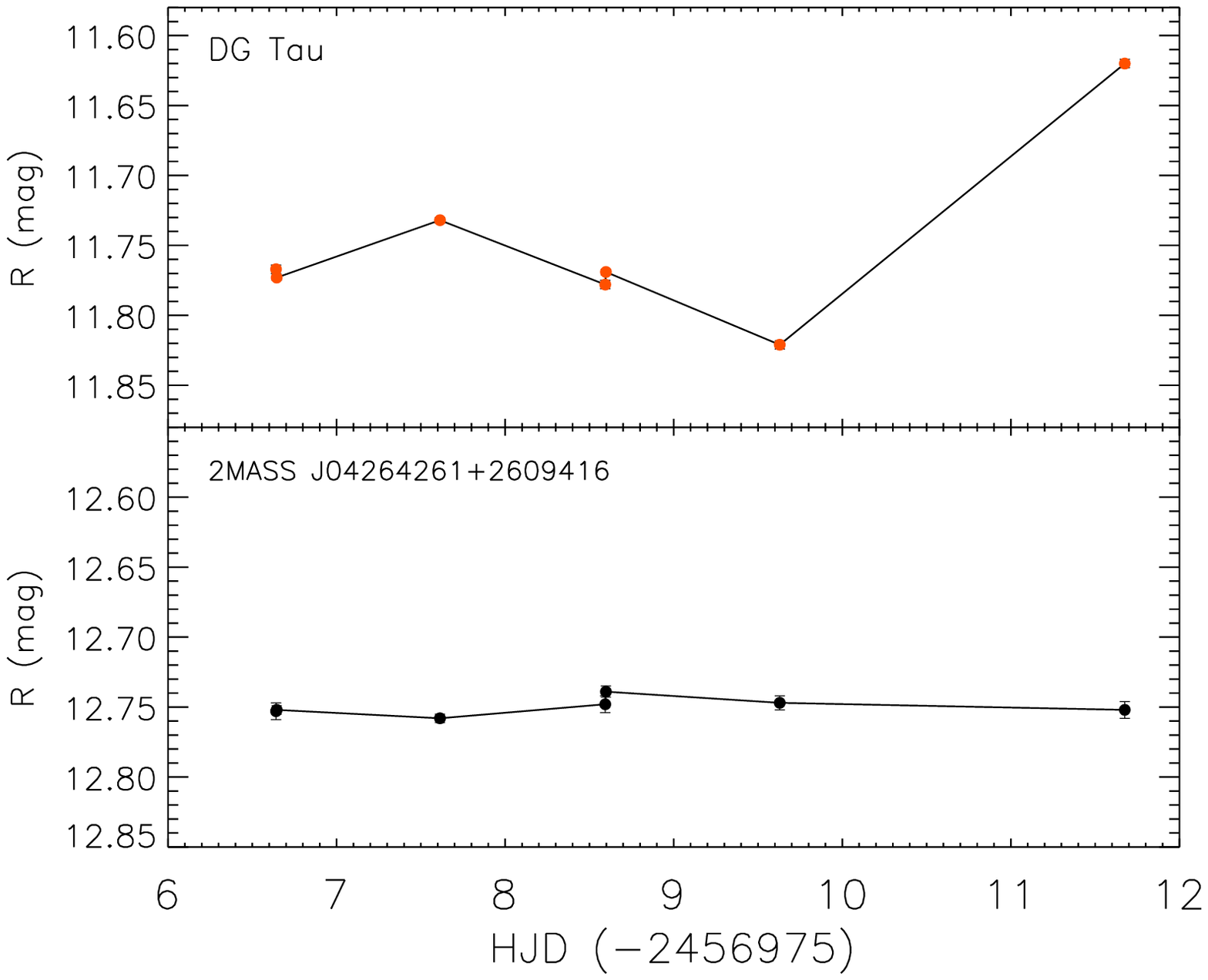}
\hspace{-0.5cm}
\includegraphics[width=6.8cm,height=6cm]{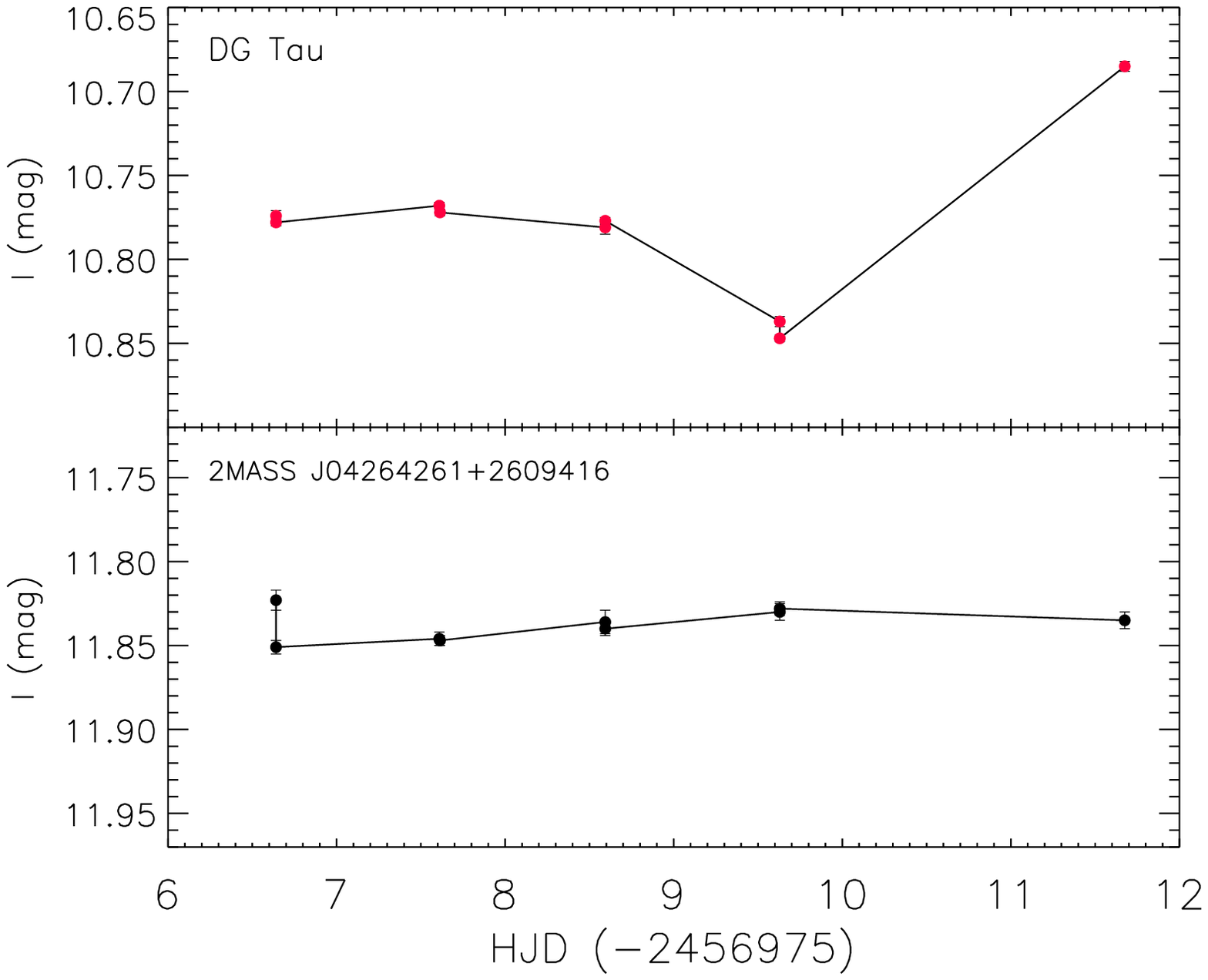}
\FigCap{$R$ and $I$ light curves of DG~Tau along with the contemporaneous photometry of a non-variable
star (2MASS\,J04264261+2609416) in the same field of view. }
\label{fig:var-comp}
\end{figure}

For the broad-band photometry, standard stars in the cluster NGC 7790 (Stetson 2000) were observed to calculate the transformation 
coefficients to the Johnson-Cousins system and the zero points in the $B$, $V$, $R$, $I$ bands. 
Local standard stars in the same fields of the targets or near them were also used to derive the zero points. 
The errors of the broad-band magnitudes include (summed in quadrature) the
photon-noise errors and the zero-point errors. 

As regards the narrow-band  photometry, we have measured the magnitude difference $CI_{\rm H\alpha}$\,=\,$H\alpha_{18}$--$H\alpha_9$, 
which is related to the intensity of the H$\alpha$ emission (see Sect.\,3.1). This index is analogous to the ($r'-H\alpha$)
color used in the IPHAS survey (Drew et al. 2005). Whenever we acquired exposures in the H$\alpha$ 
continuum ($H\alpha_{18}$) before and after the  $H\alpha_9$ frame, we have interpolated the $H\alpha_{18}$ magnitude at the time of the  
$H\alpha_9$ exposure to calculate $CI_{\rm H\alpha}$. The error of $CI_{\rm H\alpha}$ was calculated, per each star in the
CCD field, by the quadratic sum of the photon-noise errors in the two bands. When we have two  $H\alpha_{18}$ exposures before and after the 
$H\alpha_9$ one we also evaluated the standard deviation of the $H\alpha_{18}$ instrumental magnitudes and took this value as a robust 
error estimate whenever it was larger than the previous one.
The measure of $CI_{\rm H\alpha}$ for the non variable stars in the field of our targets allows us to have an independent estimate
of the precision of our measurements, which turns out to be of about 0.02\,mag for stars with a brightness comparable to our targets
(see Fig.~3 for an example).

\begin{figure}[htb]
\includegraphics[width=6.8cm,height=6cm]{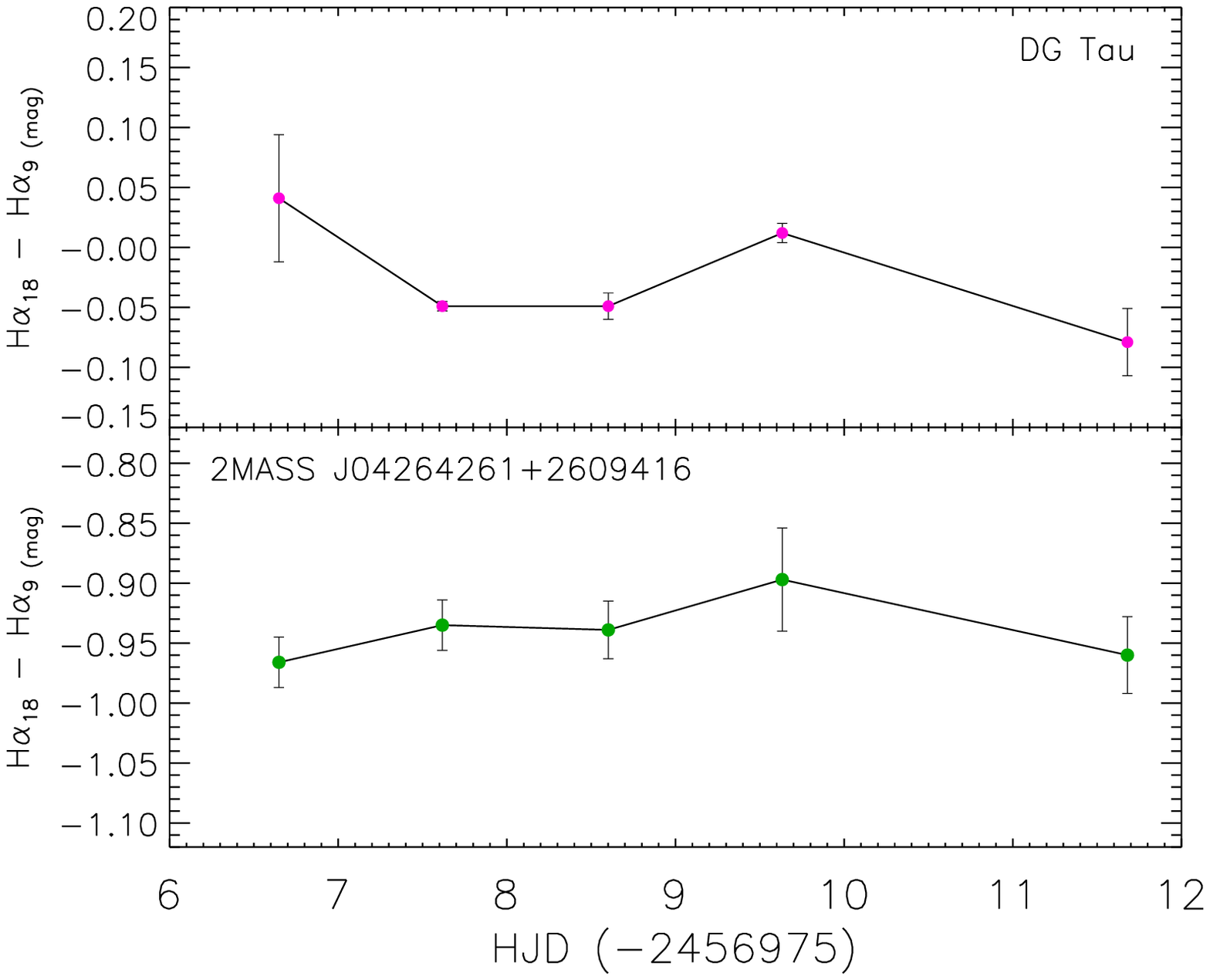}
\FigCap{H$\alpha$ index for DG~Tau and the non-variable star 2MASS\,J04264261+2609416 as a function of time. }
\end{figure}

\subsection{Spectroscopy}
\label{Sec:Spectroscopy}

Spectroscopic observations were conducted at the 2.5-m Isaac Newton Telescope (INT) with the Intermediate Dispersion Spectrograph (IDS)
during three nights, on November 14 and December 6--7, 2014. We collected at least two spectra per target during this observing run.
We used the 235-mm camera with the RED+2 CCD detector and the H1800V grating, which provides a dispersion of about 0.34~\AA\,pix$^{-1}$.
The slit width was 0.220\,mm, which corresponds to about 1.2 arcsec on the sky. The spectra were centered at about 6450\,\AA\ and approximately 
cover the range 6030--6870\,\AA.
The resolution is $R\simeq$\,9200, as we verified from the FWHM of the emission lines of the calibration spectra.
The exposure times ranged from 90 to 900 sec depending on the star magnitude. The signal-to-noise ratio (S/N) per pixel at the continuum was
in the range 30--200. 
The reduction of the spectra was made with IRAF\footnote{IRAF is distributed by the National Optical Astronomy Observatory, which is 
operated by the Association of the Universities for Research in Astronomy, inc. (AURA) under cooperative agreement with the National 
Science Foundation.}, subtracting the overscan and bias, and  flat-fielding the images. Optimal extraction of the spectra was performed
on the pre-reduced images and the spectra were wavelength calibrated using the emission lines of Cu-Ar arc spectra.
Given the small slit width (1.2") and the variable seeing (in the range 0.6"--1.6"\footnote{\tt http://www.ing.iac.es/Astronomy/development/seeing/}), we did not try to flux calibrate the spectra; we instead normalized them 
to the continuum by fitting low-order polynomials.

\section{Data analysis}
\label{sec:anal}

\subsection{Calibration of the H$\alpha$ photometry}
\label{sec:halpha}

The $H\alpha_9$ filter is centered very close to the H$\alpha$ wavelength ($\lambda_c=656.8$\,nm), while the H$\alpha_{18}$ band covers a
wavelength range that is not affected by the H$\alpha$ wings and is free from strong photospheric absorption lines (see 
Fig.~1).	
Therefore, the magnitude difference in these two bands, $CI_{\rm H\alpha}$\,=\,$H\alpha_{18}-H\alpha_9$, is an index sensitive to the intensity of
the H$\alpha$ emission. 

\begin{figure}[htb]
\includegraphics[width=9cm]{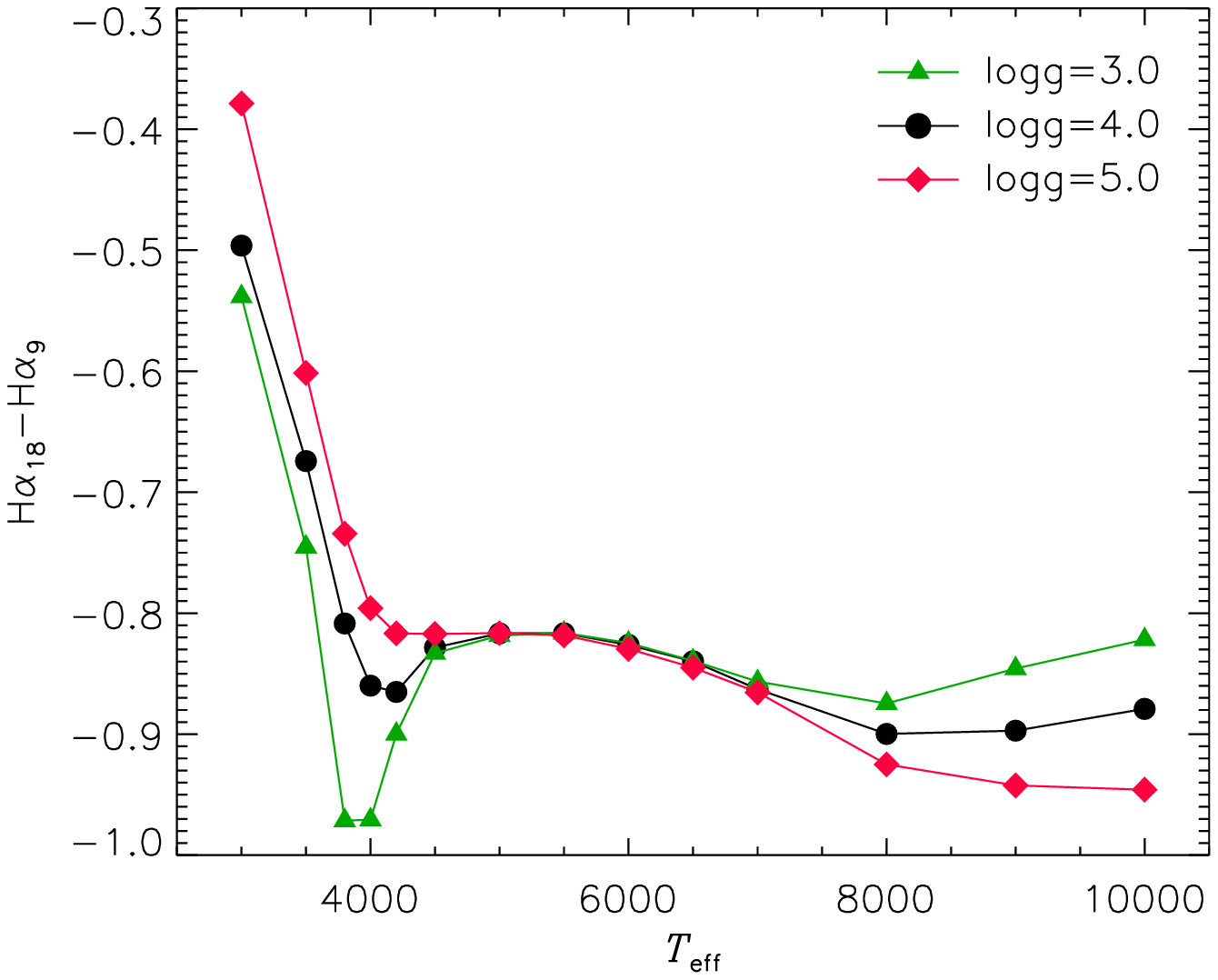}
\FigCap{Dependence of $CI_{\rm H\alpha}$ on \teff\ and \logg\  for stars without H$\alpha$ emission, as found from synthetic
photometry based on BT-Settl spectra. }
\end{figure}

However, also in absence of H$\alpha$ emission, this index depends on other atmospheric parameters, like 
\teff\ and \logg, because spectral features included in one or both H$\alpha$ bands are sensitive to them (e.g. the wings of the H$\alpha$ line
for hot stars or molecular bands for cool stars).
To investigate the dependency of $CI_{\rm H\alpha}$ on these parameters we have computed synthetic photometry based on BT-Settl spectra
(Allard et al. 2012). The synthetic magnitudes have been calculated as:
\begin{eqnarray}
H\alpha_9 & = & \int F(\lambda)R_9(\lambda)d\lambda  \\  \nonumber
H\alpha_{18} & = & \int F(\lambda)R_{18}(\lambda)d\lambda,
\end{eqnarray}
\noindent{where $R_9(\lambda)$ and $R_{18}(\lambda)$ are the response functions of the
two H$\alpha$ filters and $F(\lambda)$ is the energy flux distribution of a BT-Settl spectrum (Fig.~1).}	
The results of this synthetic photometry are shown in Fig.~\,4, which displays a weak dependency (0.05 mag peak-to peak) 
on \teff\ and negligible variations with \logg\  for FGK stars (4500\,K$\leq$\teff$\leq$7000\,K). The index $CI_{\rm H\alpha}$ becomes much more 
sensitive to \logg\ for stars hotter than 7000\,K due to the increasing dependency of the H$\alpha$ wings on surface gravity of A-type stars.
The strong \teff\ dependence for models cooler than 4500\,K is the result of TiO and CaH molecular bands.

\begin{figure}[htb]
\includegraphics[width=9cm]{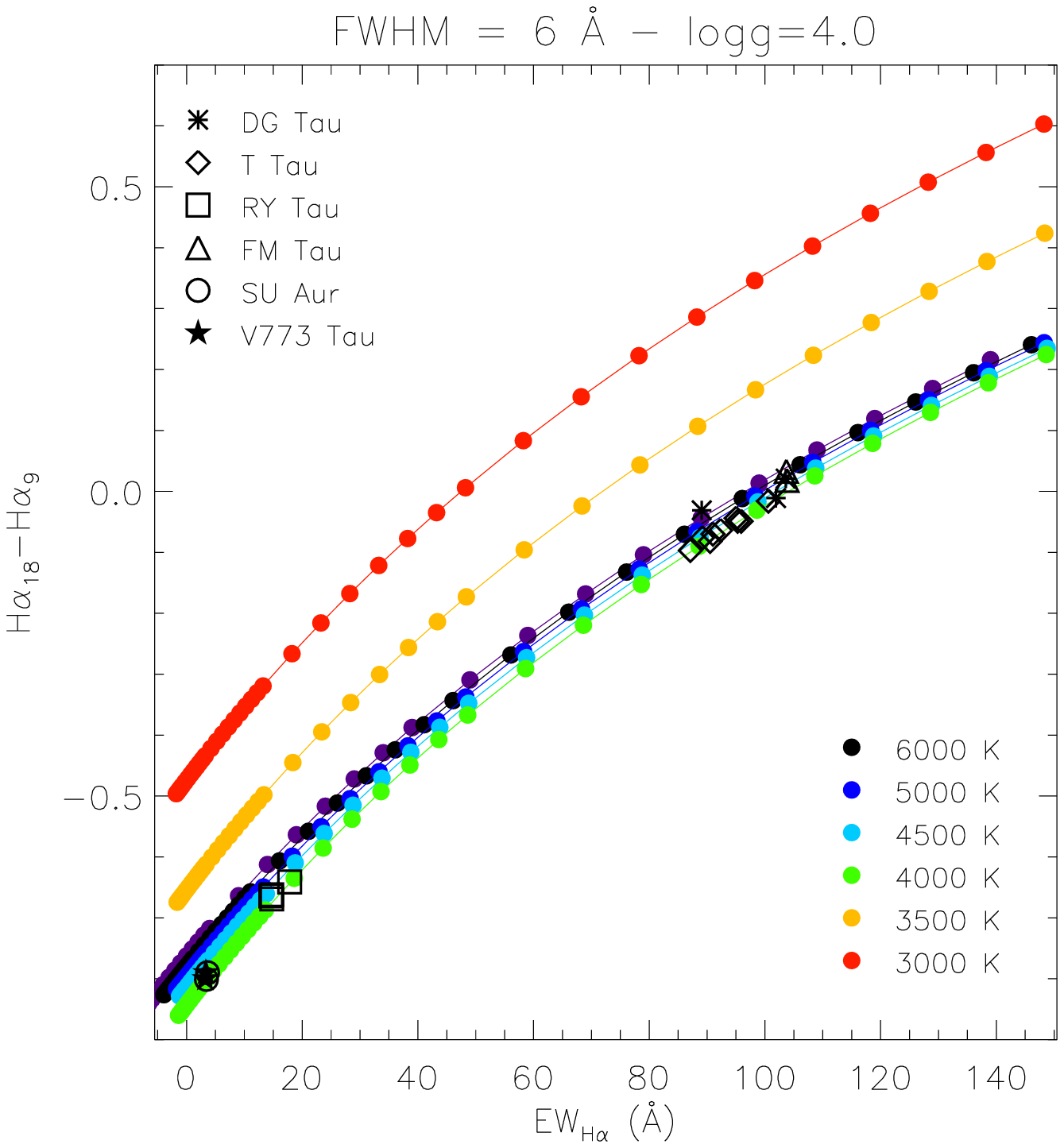}
\FigCap{Calibration relations between the quantity $CI_{\rm H\alpha}=H\alpha_{18}-H\alpha_9$ and the equivalent width, $EW_{\rm H\alpha}$.
A different colored dot corresponds to a different value of \teff, as indicated in the legend.
The values of $CI_{\rm H\alpha}$ and $EW_{\rm H\alpha}$ measured on the IDS spectra of the targeted T~Tau stars (Table~2) 
are overplotted with different symbols.}
\end{figure}

To calibrate the index $CI_{\rm H\alpha}$ versus the equivalent width of the H$\alpha$ line, $EW_{\rm H\alpha}$, we have generated synthetic 
templates with H$\alpha$ emission mimicking T\,Tauri stars.	
To this purpose, we have added Gaussian emission profiles centered at 6563\,\AA\ with increasing intensity to a BT-Settl spectrum of a 
given effective temperature (see Fig.~1 for an example). The FWHM of the Gaussian was fixed to 6\,\AA\ (274\,\kms), 
which corresponds to a full width at 10\,\%  of the maximum of about 500\,\kms\ that is a typical value for the strong accreting CTTs. 
However, the 90-\AA\  passband of the H$\alpha$ filter makes it almost insensitive to the FWHM of the emission profile, as we have 
verified with several tests.
The calibrations $EW_{\rm H\alpha}$--$CI_{\rm H\alpha}$, in the  \teff\  range 3000--6000\,K, are shown in Fig.~5.
We made these calibrations also for different values of the gravity, \logg, finding that the relations for FGK stars are basically insensitive 
to \logg, as expected on the basis of the results shown to Fig.~4.	
We note that the calibration does not change appreciably for \teff\  in the range 4000--6000\,K, while the curves at \teff=3500\,K and 3000\,K
are offset by about +0.2 and +0.4 mag, respectively. As discussed above, this \teff\ dependency is caused by molecular bands encompassed by the 
$H\alpha_{18}$ filter for low values of \teff.  This means that, if we want to convert a measure of $CI_{\rm H\alpha}$ into $EW_{\rm H\alpha}$, 
we need an independent estimate of \teff\ that should be accurate for cool objects (\teff$<4500$\,K).

A key parameter is how sensitive the index $CI_{\rm H\alpha}$ is to $EW_{\rm H\alpha}$. This can be expressed as the derivative 
of the calibration relations. We find that, for all \teff\ values, the index $CI_{\rm H\alpha}$ changes by about 0.01\,mag for a variation of 
equivalent width $\Delta EW_{\rm H\alpha}=$\,1\,\AA,  for $EW_{\rm H\alpha}<20$\,\AA, while its variation decreases to $\approx$\,0.005 
mag/\AA\  for $EW_{\rm H\alpha}\approx100$\,\AA. With the typical accuracy on $CI_{\rm H\alpha}$ of 0.02\,mag (Sect.~2.1),	
we should be able to detect  $EW_{\rm H\alpha}$ variations at the level of 2--3\,\AA.

\subsection{Analysis of the IDS spectra}
\label{sec:spectra_param}

\begin{figure}[htb]
\hspace{-1cm}
\includegraphics[width=8cm]{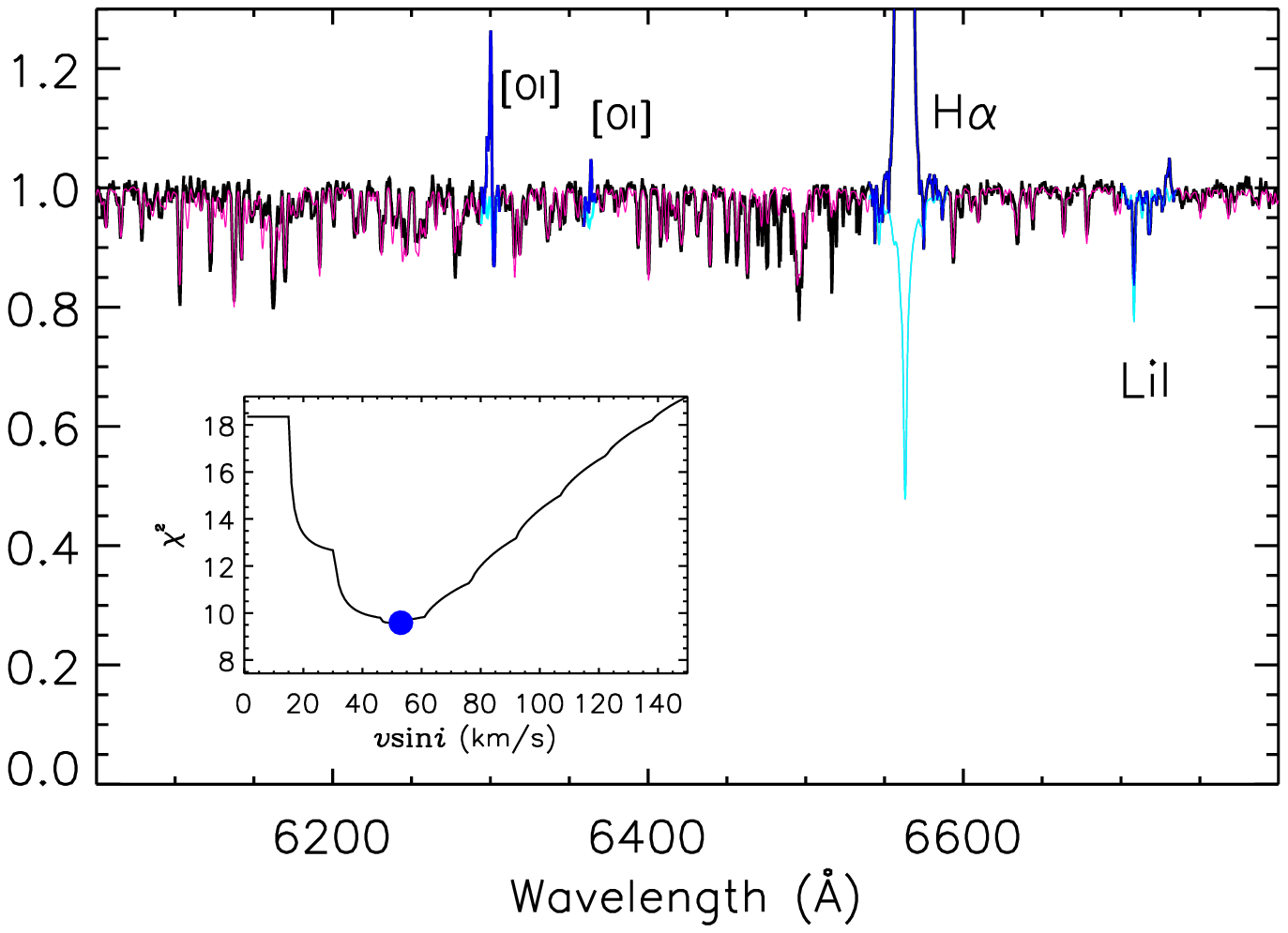}
\hspace{0cm}
\includegraphics[width=6cm,height=6cm]{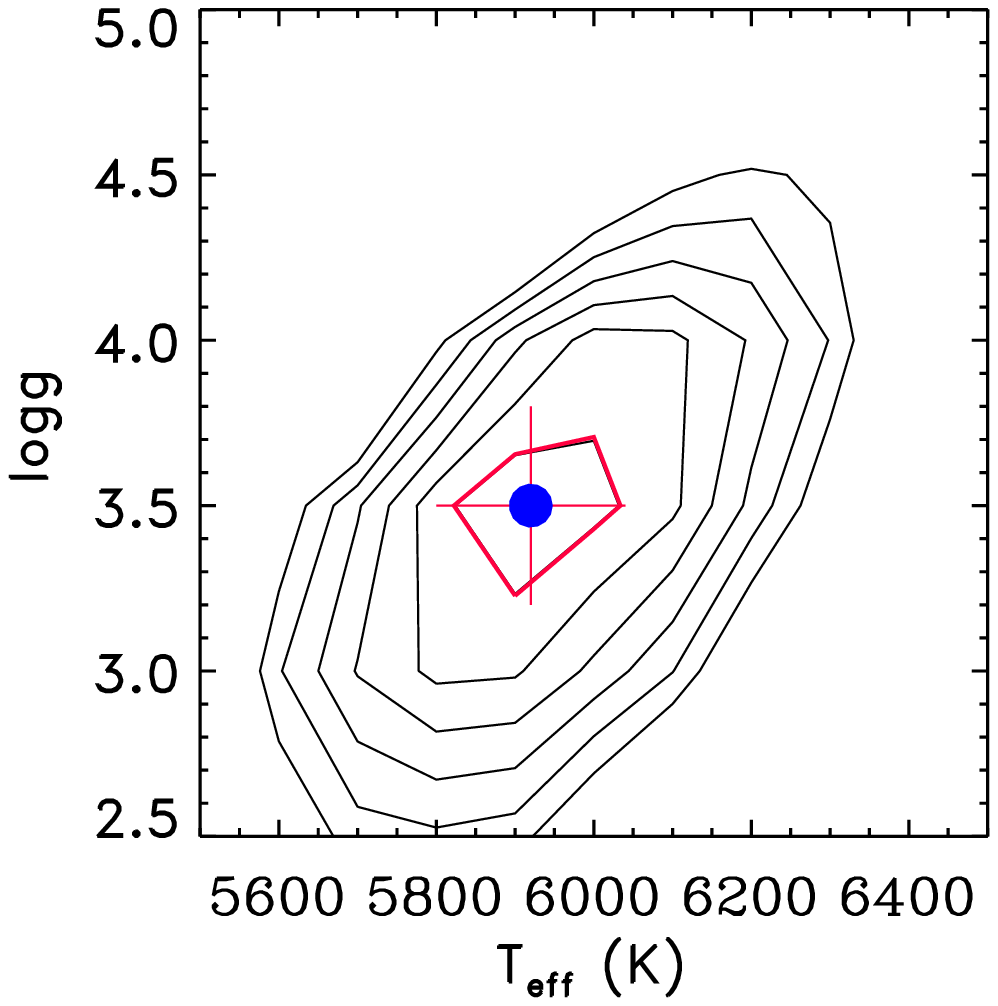}   
\FigCap{{\it Left panel:} Continuum-normalized IDS spectrum of RY~Tau (black line) with the best fitting template 
overplotted (red line). The blue and cyan lines denote the masked spectral regions of the observed and template spectrum, respectively. The inset
shows the $\chi^2$ as a function of \vsini, which clearly shows the effect of the spectral sampling ($\approx$\,15\,\kms).
{\it Right panel:} $\chi^2$ contour maps in the \teff--\logg\  plane. The blue dot indicates the best values of \teff\  and \logg, while the 1$\sigma$ 
confidence level is denoted by the thick red contour. The errorbars on \teff\  and \logg\ are also indicated.  }
\end{figure}

The IDS spectra cover the wavelength range 6000--7000\,\AA\ almost entirely. This range includes useful diagnostics of chromospheric activity, 
magnetospheric accretion and jets (e.g., [O\,{\sc i}]\,$\lambda$6300, $\lambda$6364, H$\alpha$, He\,{\sc i}\,$\lambda$6678, [S\,{\sc ii}]\,$\lambda$6716, 
$\lambda$6731) and the lithium absorption line at 6708\,\AA.
This spectral range is also suitable to determine basic stellar parameters, such as \teff, \logg, and \vsini, thanks to the numerous photospheric absorption lines.
For this task we used the code ROTFIT, adopting a grid of synthetic BT-Settl spectra (Allard et al. 2012) as templates (see Frasca et al. 2017 for more details).
We analyzed nearly all the spectral range covered by IDS with the setup selected by us; we only excluded the low-signal edges, and masked the strongest emission 
lines and the Li\,{\sc i}\,$\lambda$\,6708\,\AA\ absorption.
As shown in Frasca et al. (2017), this spectral region is well suited for the determination of \teff, \logg, and \vsini\  for K-type and early M-type stars.

An example of the results of ROTFIT for RY~Tau is shown in Fig.~6.

As a result of the strong mass accretion, the spectra of T Tauri stars can be affected by veiling, which is better observed at near-UV or optical 
wavelengths, and sometimes also in the near-IR bands (e.g., Fischer et al. 2011, Frasca et al. 2017).
For this reason, we also left the veiling free to vary in the ROTFIT processing. 
This was accomplished by adding to the templates a wavelength-independent veiling in steps of 0.1 per each iteration.
Noticeable values of the veiling, $r_{6500}$, were detected for T~Tau, DG~Tau, and FM~Tau, which are also the objects with the strongest H$\alpha$ emission. 
The $\chi^2$ profile  as a function of $r_{6500}$ is shown in Fig.~7 for T~Tau and DG~Tau. 

The stellar parameters were derived with ROTFIT using the average of the best spectra for each target and are listed in Table~2	
along with the corresponding errors. 

We measured the equivalent width of the H$\alpha$ line, $EW_{\rm H\alpha}$, and the full width at 10\% of peak height (10\%$W_{\rm H\alpha}$), which are both 
related to the mass accretion rate.
Another useful diagnostic of accretion included in our IDS spectra is the He\,{\sc i} line $\lambda$\,6678\,\AA, whose equivalent width, 
$EW_{\rm HeI}$, has been measured after the subtraction of the synthetic templates, to
remove a nearby iron line. 
The values of these parameters for each individual spectrum are reported in Table~3 along with the accretion luminosity
and mass accretion rate, calculated as described in Sect.~4.2.	

We have also measured the photometric index $CI_{\rm H\alpha}$ on each spectrum by integrating it over the $H\alpha_{18}$ and $H\alpha_{9}$ passbands, as 
we did with the synthetic templates (Sect.~3.1).		
The $EW_{\rm H\alpha}$ and $CI_{\rm H\alpha}$ values measured on the IDS spectra are overplotted, with a different symbol for each star, to the 
$EW_{\rm H\alpha}$--$CI_{\rm H\alpha}$ calibration curves in Fig.~5.		
We note that these values are closely superimposed to the curves obtained with the synthetic spectra. 

\MakeTable{lrrrrrrr}{8cm}{Stellar parameters derived from the IDS spectra.}
{\hline
\noalign{\smallskip}
Name       	   & \teff & Err & \logg & Err & \vsini& Err & $r_{6500}$                              \\  
           	   & \multicolumn{2}{c}{(K)} & \multicolumn{2}{c}{(dex)} & \multicolumn{2}{c}{(\kms)}  \\  
\hline
\noalign{\smallskip}
  T~Tau	     &  4570  &  70   &   3.20  &  0.30 &  19 &  3 & 0.5  \\  
  RY~Tau     &  5920  &  120  &   3.50  &  0.30 &  52 &  6 & 0.0  \\
V773~Tau$^a$ &  4460  &   40  &   3.30  &  0.30 &  70 &  4 & 0.0  \\  
FM~Tau$^b$   &  4070  &  190  &   4.50  &  0.90 & ... & ...& 3.5: \\  
  DG~Tau     &  4350  &   90  &   3.10  &  0.60 &  16 & 19 & 1.7  \\   
  SU~Aur     &  5870  &  110  &   3.50  &  0.20 &  61 &  7 & 0.0  \\ 
\hline
\noalign{\smallskip}
\multicolumn{8}{p{8cm}}{$^a$ Multiple system. Uncertain stellar parameters. }\\ 
\multicolumn{8}{p{8cm}}{ $^{b}$ Uncertain stellar parameters and no \vsini\ determination because of the low S/N ratio at the continuum.}\\
}

\begin{figure}[htb]
\includegraphics[width=6.5cm]{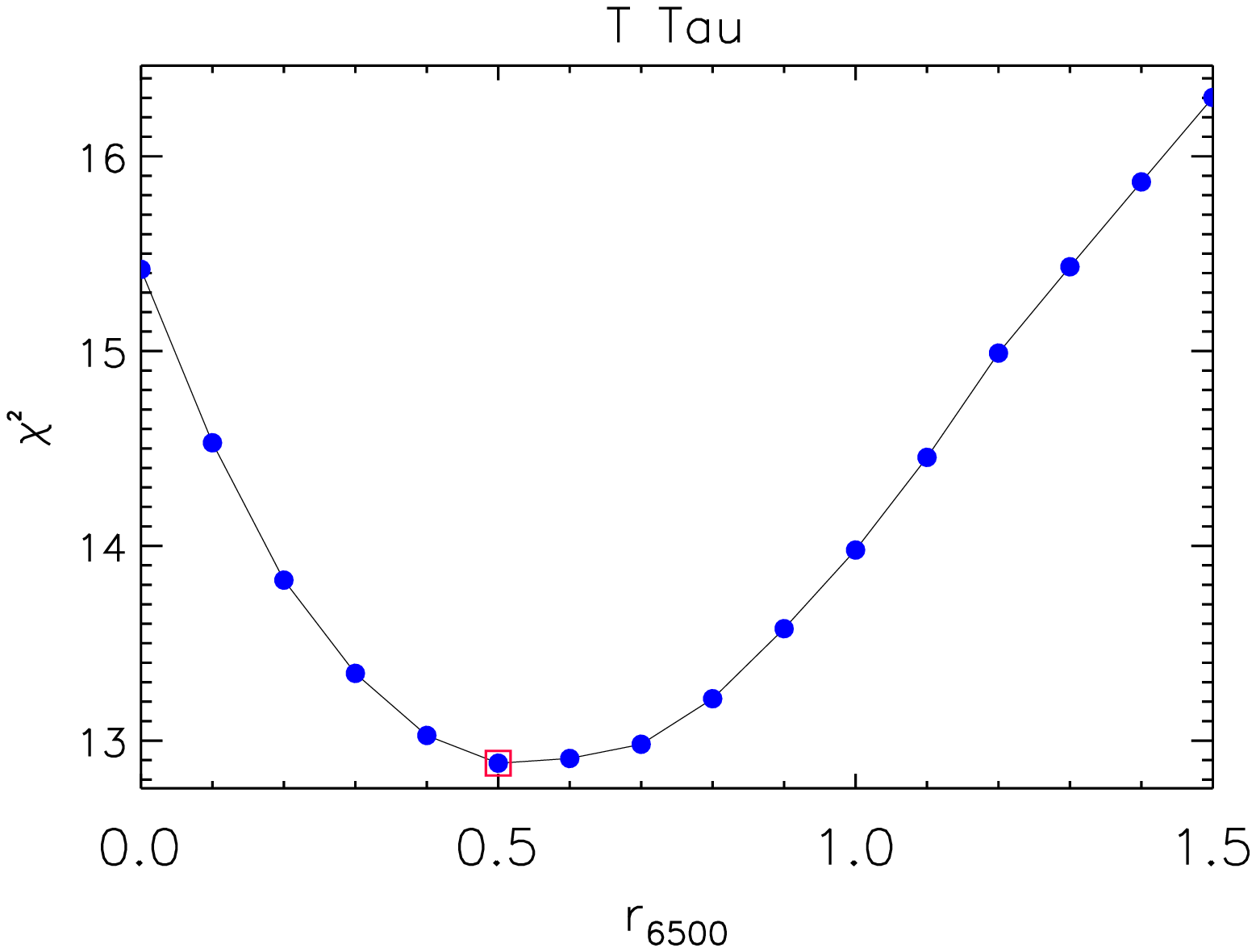}    
\includegraphics[width=6.5cm]{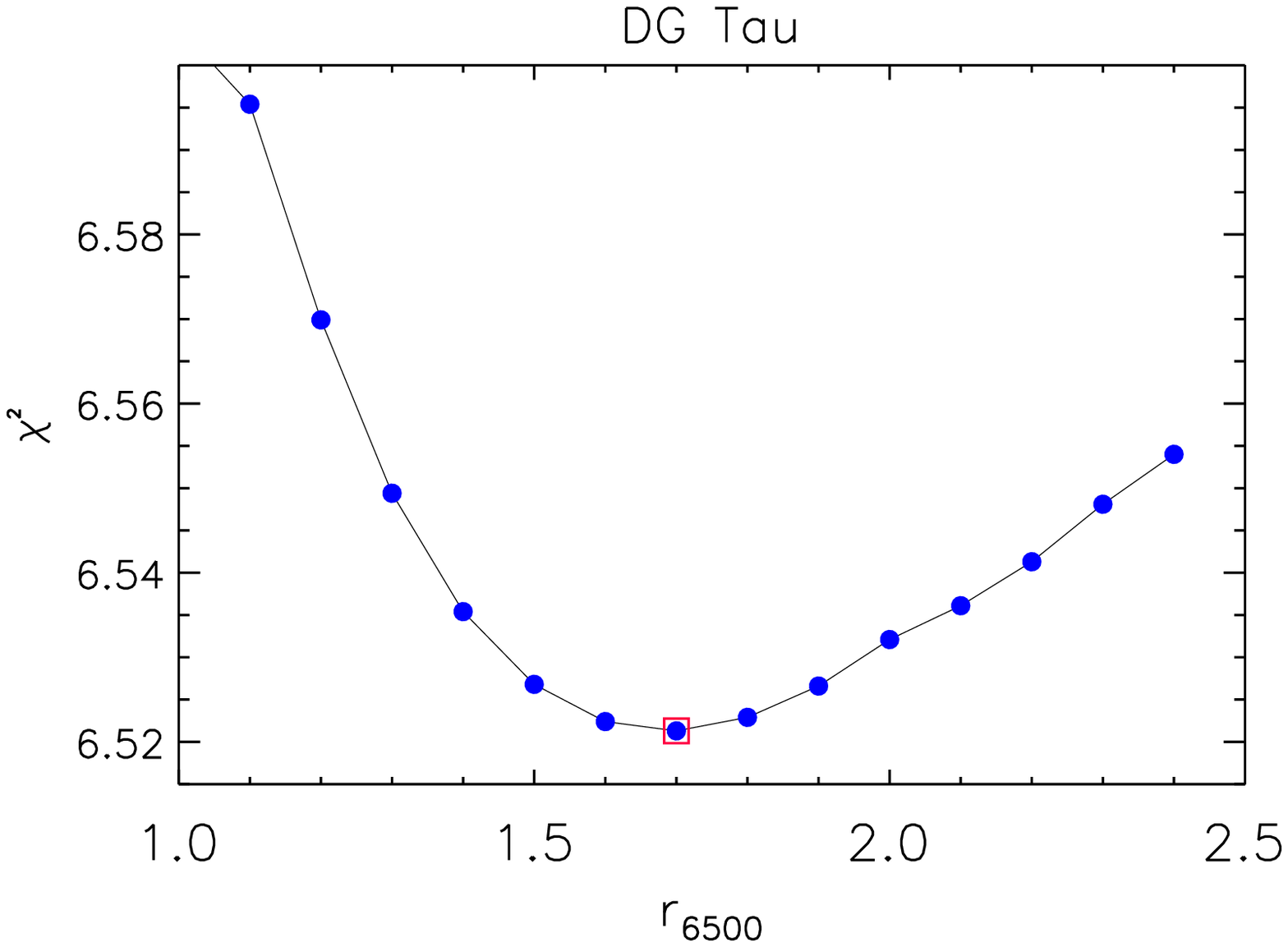}    
\FigCap{$\chi^2$ minimization of veiling for T~Tau {\it (left panel)} and DG~Tau {\it (right panel)}.}
\label{fig:veiling}
\end{figure}

\section{Results}
\label{sec:results}

\subsection{Stellar parameters}
\label{subsec:Stellar_param}

\begin{figure}[th]
\hspace{-1cm}
\includegraphics[width=8.5cm]{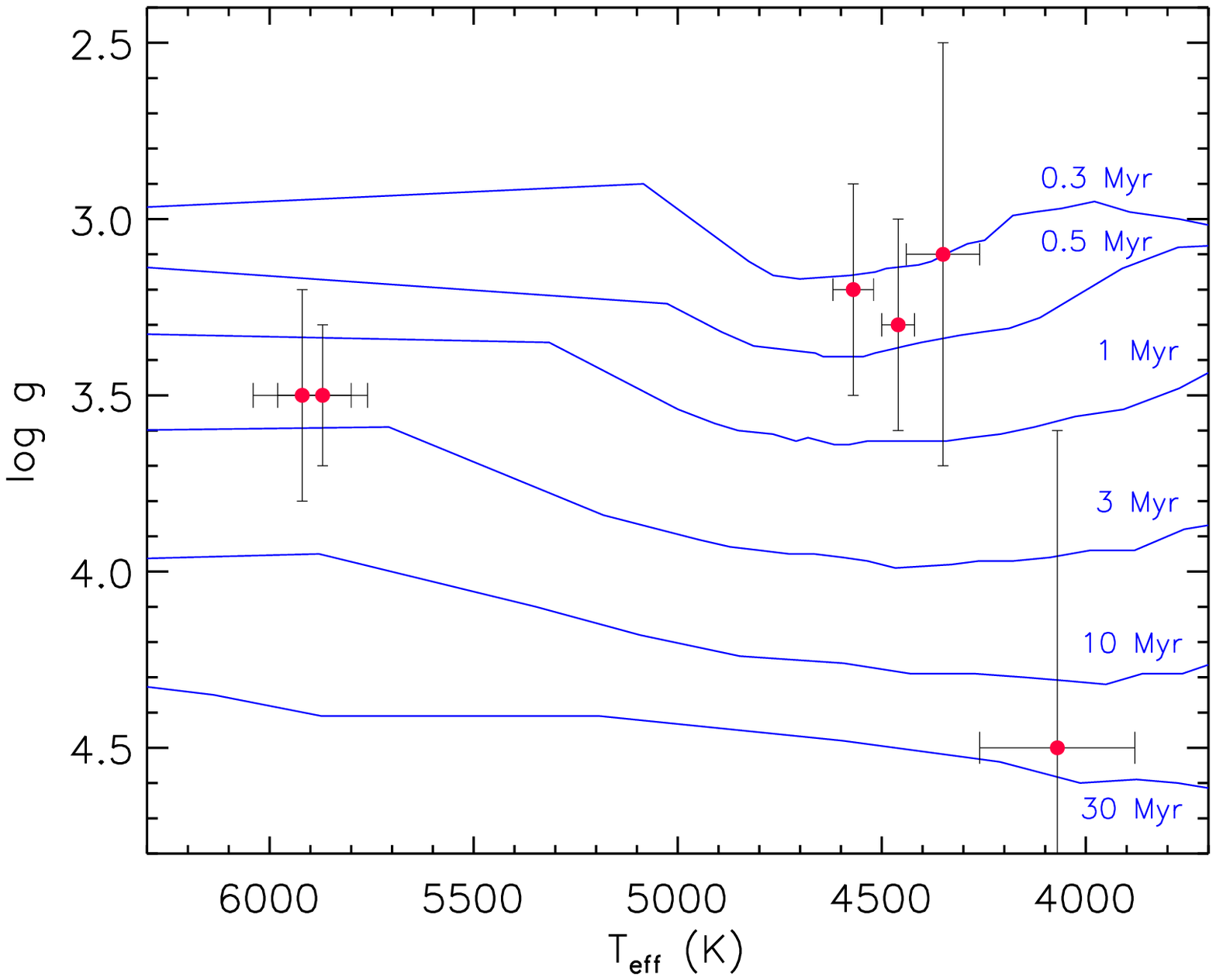}
\FigCap{$T_{\rm eff}$--$\log g$ diagram. The full blue lines are the PMS isochrones with a solar metallicity from Siess et al. (2000) at 0.3, 0.5, 1, 3, 10, and 30\,Myr.}
\label{Fig:LoggTeff}
\end{figure}

The determination of stellar parameters from the spectra of CTTs is hampered by the presence of strong emission lines and continuum emission arising
from the matter accreting onto the central star that overlaps the photospheric spectrum giving rise to the veiling.  
For this reason one can find very different determinations of effective temperature and spectral type in the literature. This can also be 
the result of the different spectral ranges and resolutions of the used spectra.
 Moreover, in some cases the parameters are not directly measured
from spectra, but they are derived from photometric indices. 
For the \vsini, instead, it is normally found a better consistency among the various values in the literature. 

As an example, among the catalogs available in the {\it VizieR Service for Astronomical Catalogues} at {\it Centre de Donn\'ees astronomiques de 
Strasbourg} (CDS), a SpType ranging from F8\,V to K1\,IV and \teff\ in the range 2600--6890\,K are listed for T~Tau. 
The values \teff=4750\,K derived by Wright et al. (2003) from a SpType--\teff\  calibration and  \teff=4817\,K reported by Huber et al. (2016) are in 
marginal agreement with our determination of \teff=4570$\pm$70\,K.  We note that the inclusion of veiling among the parameters derived with our code
simultaneously to \teff, \logg, and \vsini\  allows an unbiased determination of these parameters; the effect of the veiling is to 
fill the core of the absorption lines, mimicking a higher photospheric temperature, so that, when it is taken into account, lower \teff\ values are found.
The only determination of surface gravity available in the CDS is \logg=2.73 by Huber et al. (2016).
We note that T~Tau has two companions (T~Tau\,Sa and Sb) with a separation of about 0.7" (Dyck et al. 1982, Kasper et al. 2016) which are so weak ($\Delta J>10$\,mag,
$\Delta H\sim$\,5.5\,mag, $\Delta K\sim$\,2.2\,mag for T~Tau\,Sa, Kasper et al. 2016) that they do not affect either the determination of its 
atmospheric parameters or its optical photometry.
The values of \vsini\ of 23.0\,$\pm$\,1.2\,\kms\ (Nguyen et al. 2012), 20.1\,$\pm$\,1.2\,\kms\ (Hartmann et al. 1986), and 20.7\,\kms\ (Glebocki  \& Gnacinski 2005) 
are all in fairly good agreement with each other and with our value of 19\,$\pm$\,3\,\kms.
A low value of veiling a 7500\,\AA, $r_{7500}=0.1$, is reported by Herczeg \& Hillenbrand (2014), which is in line with our slightly larger value, $r_{6500}=0.5$, 
at a shorter wavelength. 

For RY~Tau, our determination of \teff=5920$\pm$120\,K is in very good agreement with the value of 5945\,K reported by Kim et al. (2013), which is based on 
Spitzer low-resolution IR spectroscopy. It is also in line with the spectral classification G1--2\,IV by Petrov et al. (1999) and G0 by Herczeg \& Hillenbrand (2014). 
However, very different values of \teff, ranging from 4920\,K to 5080\,K and based on SpType--\teff\ calibrations, are also reported in previous works 
(e.g., Kenyon \& Hartmann 1995, Davies et al. 2014).
Our \vsini=52$\pm$6\,\kms\ agrees very well with the literature values of 55$\pm$3\,\kms\ (Mora et al. 2001), 52$\pm$2\,\kms\ (Petrov et al. 1999), and 
48.7$\pm$3.8\,\kms\ (Hartmann et al. 1986).

The stellar parameters of V773~Tau must be taken with care, because this object is indeed a multiple system composed of three stars, separated 
by 50--110 mas (A--B) and about 250 mas (A--C) (e.g., Boden et al. 2012, and references therein). The brightest star (A) is also a double-lined spectroscopic 
binary with components of SpType K2 and K5 (Welty 1995) and an orbital period of about 51 days. 
Our value of \vsini=70\,\kms\ is larger than the values of about 41\,\kms\ reported by Welty (1995) for each component.
It is likely the effect of the blended lines of the individual components, which are not resolved with the intermediate resolution of the IDS spectrograph.

For FW~Tau the accuracy of the stellar parameters is low, mainly because of the low signal-to-noise ratio at the continuum, S/N$\approx$\,30, and the 
strong veiling. 

In addition to T~Tau, the objects with a veiling significantly larger than zero are FM~Tau and DG~Tau. Our value of $r_{6500}=3.5$ for FM~Tau is quite 
uncertain, due to the low S/N ratio of our IDS spectra. Values of $r_{7500}$ in the range 0.45--0.72 are reported by Herczeg \& Hillenbrand (2014).
For DG~Tau they quote $r_{7500}$=0.43--0.66 and Fischer et al. (2011) measured the veiling at different wavelengths with $r_{6500}\sim 1.0$, which is
smaller than the value of 1.7 measured by us. They derived larger veiling values, up to about 2.0, only at wavelengths shorter than 5000\,\AA. 
These differences can either arise from the different methods used to measure the veiling or from a real variation due to a variable 
mass accretion rate.

A noteworthy result is the determination of surface gravity that, excluding FM~Tau, is in the range \logg\,=\,3.1--3.5.
As shown in the \logg\--\teff\ diagram (Fig.~8), where the pre-main sequence (PMS) isochrones of Siess et al. (2000) are displayed
with blue lines, the position of our targets (red dots) suggests ages in the range 0.3-3\,Myr, which is in line with the 
values from the literature (e.g., Kenyon et al. 1990, Briceno et al. 1999).

\subsection{Accretion parameters}
\label{subsec:Accretion}

The availability of optical photometry nearly contemporaneous to the IDS spectra has allowed us to get an absolute flux scale for 
the IDS spectra. This permits to convert the equivalent widths of the emission lines into fluxes and luminosities.
In particular, from the spectral energy distribution in the $BVRI$ bands and the stellar parameters reported in Table~2,	
we have evaluated the extinction, $A_V$, and the flux at Earth in the $R$ band corrected for the extinction, $f_{R_0}$, which 
are listed in Table~4.	
We found for all sources values of $A_V$ larger than those reported by Herczeg \& Hilenbrabd (2014) but smaller or equal to 
those derived from near-IR analyses (e.g., Rebull et al. 2010, Furlan et al. 2011); our values agree reasonably well with those reported by LopezMartinez 
et al.~(2015). 

Then the line luminosity, \Ll, can be calculated as
\begin{equation}
L_{\rm line} = 4\pi d^2 f_{R_0} EW_{\rm line}
\end{equation}
\noindent{where $d$ is the distance to the star (Table~1).}		
We adopted the $f_{R_0}$ value from the OAC photometry closest in time to the IDS spectra. 

We used the relations between the accretion luminosity, \Lacc, and \Ll\ proposed by Alcal\'a et al. (2017, Table~B.1) to derive \Lacc\ 
from the two accretion diagnostics in the IDS spectra, namely H$\alpha$ and He\,{\sc i}\,$\lambda$\,6678\,\AA.
These two sets of \Lacc\ values are reported in Table~3.		
The \Lacc\ errors include the errors on \Ll, which are dominated by the $f_{R_0}$ error (from 7 to 30\,\%), the error of the distance 
(from 4 to 15\,\%), and the standard deviation of the \Lacc--\Ll\ relations reported by Alcal\'a et al. (2017).  
We find a marginal agreement between the \Lacc\ values derived from H$\alpha$ and He\,{\sc i} lines, with differences of 0.5\,dex, at most.
These differences can be ascribed to both the low intensity of the He\,{\sc i} line, which may yield less accurate results, and to the contribution
of different phenomena giving rise to the H$\alpha$ emission.
It has been shown, indeed, that in accreting stars the Balmer emission often displays a relevant optically-thin component which arises from the accretion 
funnels and overlaps the emission produced by the accretion shock over the stellar photosphere (e.g., Alcal\'a et al. 2017, Frasca et al. 2017, and references therein).

The accretion luminosities were converted into mass accretion rates, \Macc, using the relation
\begin{equation}
\dot{M}_{acc} = ( 1 - \frac{R_{\star}}{R_{\rm in}} )^{-1} ~ \frac{L_{acc} R_{\star}}{G M_{\star}}
 \approx 1.25 ~ \frac{L_{acc} R_{\star}}{G M_{\star}} 
\end{equation}
\noindent
where $R_{\star}$ and $R_{\rm in}$ are the stellar radius and inner-disc radius, respectively, and \Mstar\ is the star mass
(Gullbring et al. 1998, Hartmann et al. 1998).
Here we assumed  $R_{\rm in}$ to be $5\,R_\star$, as in Alcal\'a et al. 2017 and previous studies.
The $R_{\star}$ and \Mstar\ values were taken from Herczeg \& Hillenbrand (2014) and are also reported in Table~1.	
The \Macc\ values, calculated both from H$\alpha$ and He\,{\sc i}, are also listed in Table~3.		
We estimate an average uncertainty in \Macc$^{H\alpha}$\ of about 0.5\,dex. 

\setlength{\tabcolsep}{1pt}

\MakeTable{lllrrrccrrcc}{16cm}{Equivalent widths, $H\alpha$ color index, and accretion parameters derived from the IDS spectra.}
{\hline
\noalign{\smallskip}
Name       	  & HJD          & $10\%W_{\rm H\alpha}$ & $EW_{\rm H\alpha}$~~~ & $CI_{\rm H\alpha}$ & $EW_{\rm HeI}$~~~ & $A_V$ & $f_{R_0}$	& $\log$\Lacc$^{H\alpha}$ & $\log$\Lacc$^{HeI}$ & {\scriptsize $\log$\Macc$^{H\alpha}$} & {\scriptsize $\log$\Macc$^{HeI}$} \\  
           	  & 2\,456\,900+ &  (\kms)               & (\AA)~~~              & (mag)	      & (\AA)~~~	  & (mag) & * & (\Lsun)		  & (\Lsun)		& {\scriptsize (\Msun\,yr$^{-1}$)}	 & {\scriptsize (\Msun\,yr$^{-1}$)} \\  
\hline
\noalign{\smallskip}
  T~Tau	   &  76.7366 &  482$\pm$23  &   100.60$\pm$0.19  &   $-0.016$   & 0.28$\pm$0.04 & 1.7  & 1.28(9)e-12 & $ 0.49\pm0.46$ & $ 0.12\pm0.46$  &  $-6.64$  &  $-7.01$ \\    
  T~Tau	   &  98.4723 &  377$\pm$ 8  &    95.89$\pm$0.16  &   $-0.049$   & 0.19$\pm$0.05 & 1.6  & 1.18(8)e-12 & $ 0.42\pm0.46$ & $-0.14\pm0.52$  &  $-6.70$  &  $-7.26$ \\ 
  T~Tau	   &  98.4761 &  376$\pm$15  &    95.03$\pm$0.23  &   $-0.044$   & 0.24$\pm$0.08 & 1.6  & 1.18(8)e-12 & $ 0.42\pm0.46$ & $-0.01\pm0.56$  &  $-6.71$  &  $-7.14$ \\  
  T~Tau	   &  98.4790 &  377$\pm$15  &    95.27$\pm$0.28  &   $-0.049$   & 0.24$\pm$0.09 & 1.6  & 1.18(8)e-12 & $ 0.42\pm0.46$ & $-0.01\pm0.58$  &  $-6.71$  &  $-7.14$ \\  
  T~Tau	   &  98.7006 &  392$\pm$ 8  &    92.33$\pm$0.32  &   $-0.064$   & 0.11$\pm$0.03 & 1.6  & 1.18(8)e-12 & $ 0.40\pm0.46$ & $-0.43\pm0.48$  &  $-6.72$  &  $-7.56$ \\  
  T~Tau	   &  98.7057 &  390$\pm$ 8  &    89.22$\pm$0.54  &   $-0.076$   & 0.15$\pm$0.03 & 1.6  & 1.18(8)e-12 & $ 0.39\pm0.46$ & $-0.26\pm0.49$  &  $-6.74$  &  $-7.39$ \\  
  T~Tau	   &  98.7088 &  377$\pm$22  &    87.11$\pm$0.22  &   $-0.097$   & 0.19$\pm$0.05 & 1.6  & 1.18(8)e-12 & $ 0.38\pm0.46$ & $-0.14\pm0.52$  &  $-6.75$  &  $-7.26$ \\
  T~Tau	   &  99.6538 &  392$\pm$ 8  &    90.53$\pm$0.29  &   $-0.082$   & 0.14$\pm$0.02 & 1.6  & 1.18(8)e-12 & $ 0.39\pm0.46$ & $-0.30\pm0.46$  &  $-6.73$  &  $-7.43$ \\  
  T~Tau	   &  99.6594 &  390$\pm$45  &    91.01$\pm$0.26  &   $-0.070$   & 0.11$\pm$0.03 & 1.6  & 1.18(8)e-12 & $ 0.40\pm0.46$ & $-0.43\pm0.52$  &  $-6.73$  &  $-7.56$ \\  
  RY~Tau   &  76.7259 &  602$\pm$ 8  &    17.84$\pm$0.12  &   $-0.641$   & 0.09$\pm$0.02 & 2.3  & 1.8(5)e-12  & $ 0.05\pm0.63$ & $-0.04\pm0.64$  &  $-7.18$  &  $-7.27$ \\    
  RY~Tau   &  98.5170 &  633$\pm$15  &    14.74$\pm$0.36  &   $-0.669$   & 0.10$\pm$0.02 & 2.4  & 2.1(6)e-12  & $ 0.03\pm0.63$ & $ 0.09\pm0.64$  &  $-7.21$  &  $-7.14$ \\    
  RY~Tau   &  98.5218 &  619$\pm$15  &    14.57$\pm$0.38  &   $-0.664$   & 0.05$\pm$0.04 & 2.4  & 2.1(6)e-12  & $ 0.02\pm0.63$ & $-0.28\pm1.18$  &  $-7.21$  &  $-7.52$ \\   
  RY~Tau   &  98.5297 &  634$\pm$15  &    14.74$\pm$0.28  &   $-0.662$   & 0.10$\pm$0.01 & 2.4  & 2.1(6)e-12  & $ 0.03\pm0.63$ & $ 0.09\pm0.62$  &  $-7.21$  &  $-7.14$ \\   
  V773~Tau &  99.6959 &  393$\pm$15  &     3.24$\pm$0.10  &   $-0.800$   & 0.04$\pm$0.02 & 2.9  & 5.0(3)e-13  & $-1.67\pm0.48$ & $-1.46\pm0.67$  &  $-8.58$  &  $-8.37$ \\    
  V773~Tau &  99.7085 &  410$\pm$15  &     3.33$\pm$0.09  &   $-0.790$   & 0.08$\pm$0.02 & 2.9  & 5.0(3)e-13  & $-1.66\pm0.48$ & $-1.08\pm0.52$  &  $-8.57$  &  $-7.99$ \\  
  FM~Tau   &  99.6959 &  376$\pm$ 8  &   103.81$\pm$0.66  &   $ 0.018$   & 1.53$\pm$0.20 & 0.7  & 1.20(7)e-14 & $-1.80\pm0.48$ & $-1.50\pm0.47$  &  $-8.39$  &  $-8.09$ \\  
  FM~Tau   &  99.7085 &  391$\pm$15  &   103.70$\pm$0.88  &   $ 0.033$   & 1.69$\pm$0.18 & 0.7  & 1.20(7)e-14 & $-1.80\pm0.48$ & $-1.45\pm0.46$  &  $-8.39$  &  $-8.04$ \\  
  DG~Tau   &  76.7318 &  512$\pm$15  &   101.91$\pm$0.39  &   $-0.011$   & 0.76$\pm$0.09 & 2.4  & 3.0(4)e-13  & $-0.21\pm0.57$ & $-0.12\pm0.55$  &  $-7.29$  &  $-7.19$ \\  
  DG~Tau   &  98.6024 &  499$\pm$15  &    89.07$\pm$1.26  &   $-0.031$   & 1.20$\pm$0.28 & 2.1  & 2.7(4)e-13  & $-0.32\pm0.57$ & $ 0.08\pm0.58$  &  $-7.40$  &  $-6.99$ \\   
  DG~Tau   &  98.6077 &  515$\pm$ 9  &   103.60$\pm$0.51  &   $ 0.021$   & 0.95$\pm$0.20 & 2.1  & 2.7(4)e-13  & $-0.25\pm0.57$ & $-0.04\pm0.57$  &  $-7.32$  &  $-7.12$ \\ 
  SU~Aur   &  76.7415 &  527$\pm$15  &     3.44$\pm$0.06  &   $-0.801$   & 0.07$\pm$0.01 & 2.4  & 2.4(4)e-12  & $-0.84\pm0.53$ & $-0.27\pm0.52$  &  $-8.12$  &  $-7.55$ \\  
  SU~Aur   &  98.5475 &  528$\pm$15  &     3.64$\pm$0.16  &   $-0.790$   & 0.01$\pm$0.01 & 1.8  & 2.3(3)e-12  & $-0.83\pm0.52$ & \dots~~~~~~     &  $-8.11$  &  ~~\dots \\  
\hline
\noalign{\smallskip}
\multicolumn{12}{p{16cm}}{* = (\fluxunit).}\\
}

\normalsize

Our values of \Lacc\ and \Macc\ agree reasonably well with the literature values, with the exception of T~Tau for which we found values of $\log$\Macc$^{H\alpha}\approx$\,0.5--0.8 dex larger 
than those from the literature. The accretion rates we derive for T~Tau from the He\,{\sc i}\,$\lambda$6678 line are instead in good agreement with the values of $\log$\Macc\ of $[-7.5,-7.2]$ 
and $-7.34$ reported by Calvet et al. (2004) and White \& Ghez (2001), respectively. For RY~Tau, the mass accretion rates we derive are in agreement 
with the range of values  $[-7.2,-7.0]$ quoted by Calvet et al. (2004). 
The same holds for SU~Aur, for which Calvet et al. (2004) report $\log$\Macc$\simeq-8.2$, and for DG~Tau ($\log$\Macc$\simeq-7.34$, White \& Ghez 2001). 
The value of $\log$\Macc$^{H\alpha}\simeq-8.4$ that we found for FM~Tau is in between the values of $-8.1$ and $-8.9$ reported by Valenti et al. (1993) and White \& Ghez (2001), respectively.
 
We point out that, using the calibrations between $CI_{\rm H\alpha}$ and $EW_{\rm H\alpha}$ of Sect.~3.1 and the simultaneous broad-band photometry, we can calculate
\Lacc\ from the photometric data alone, if the distance to the object is known. As an example we show in Fig.~9 the accretion luminosity as a function of time for T~Tau. 

\begin{figure}[th]
\hspace{0cm}
\includegraphics[width=7cm]{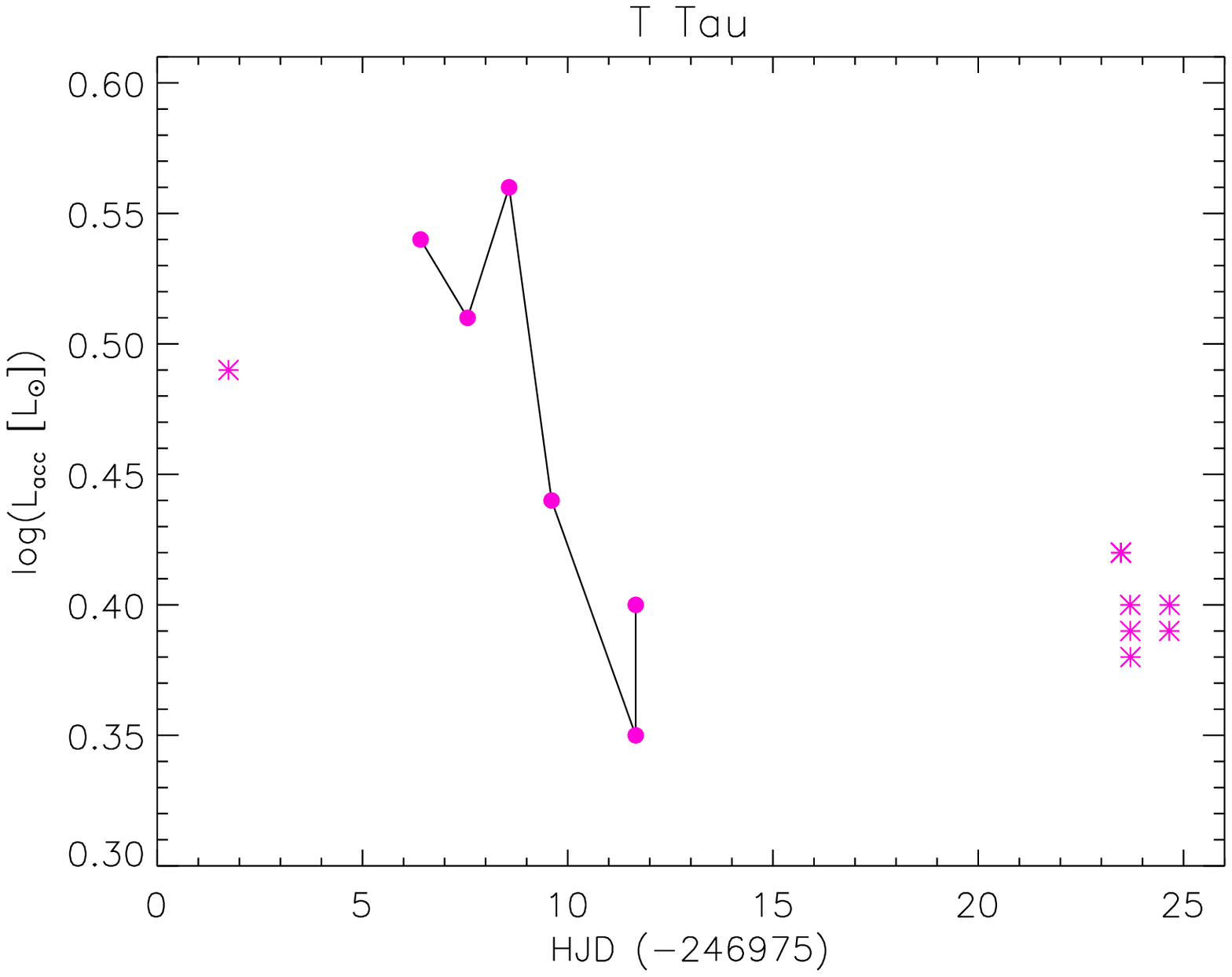}
\FigCap{Accretion luminosity for T~Tau calculated from IDS spectra (asterisks) and from OAC photometry (dots) as a function of time.}
\end{figure}

\subsection{Broad-band and H$\alpha$ variations}

\begin{figure}[htb]
\includegraphics[width=4.3cm]{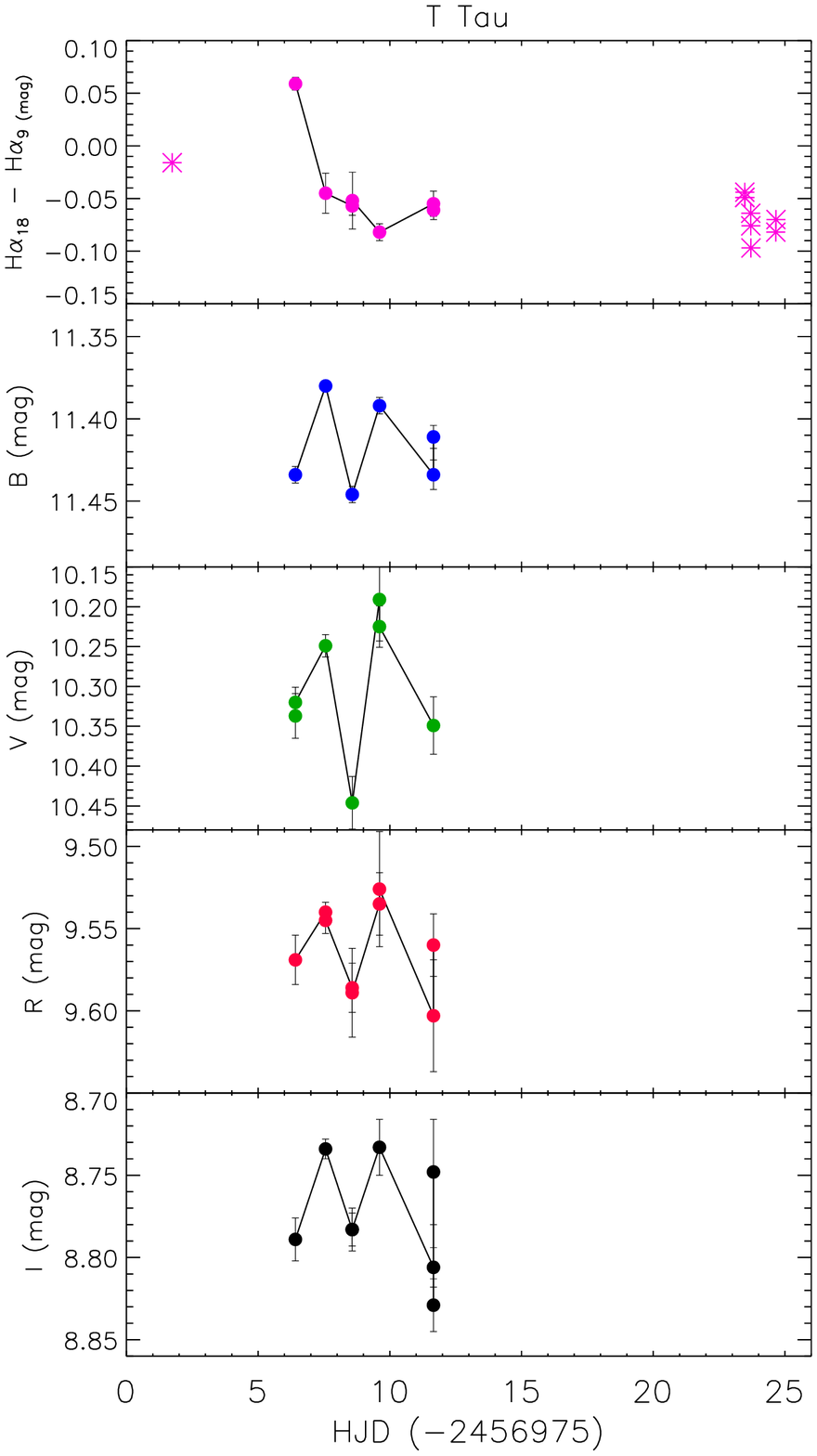}
\hspace{-0.4cm}
\includegraphics[width=4.3cm]{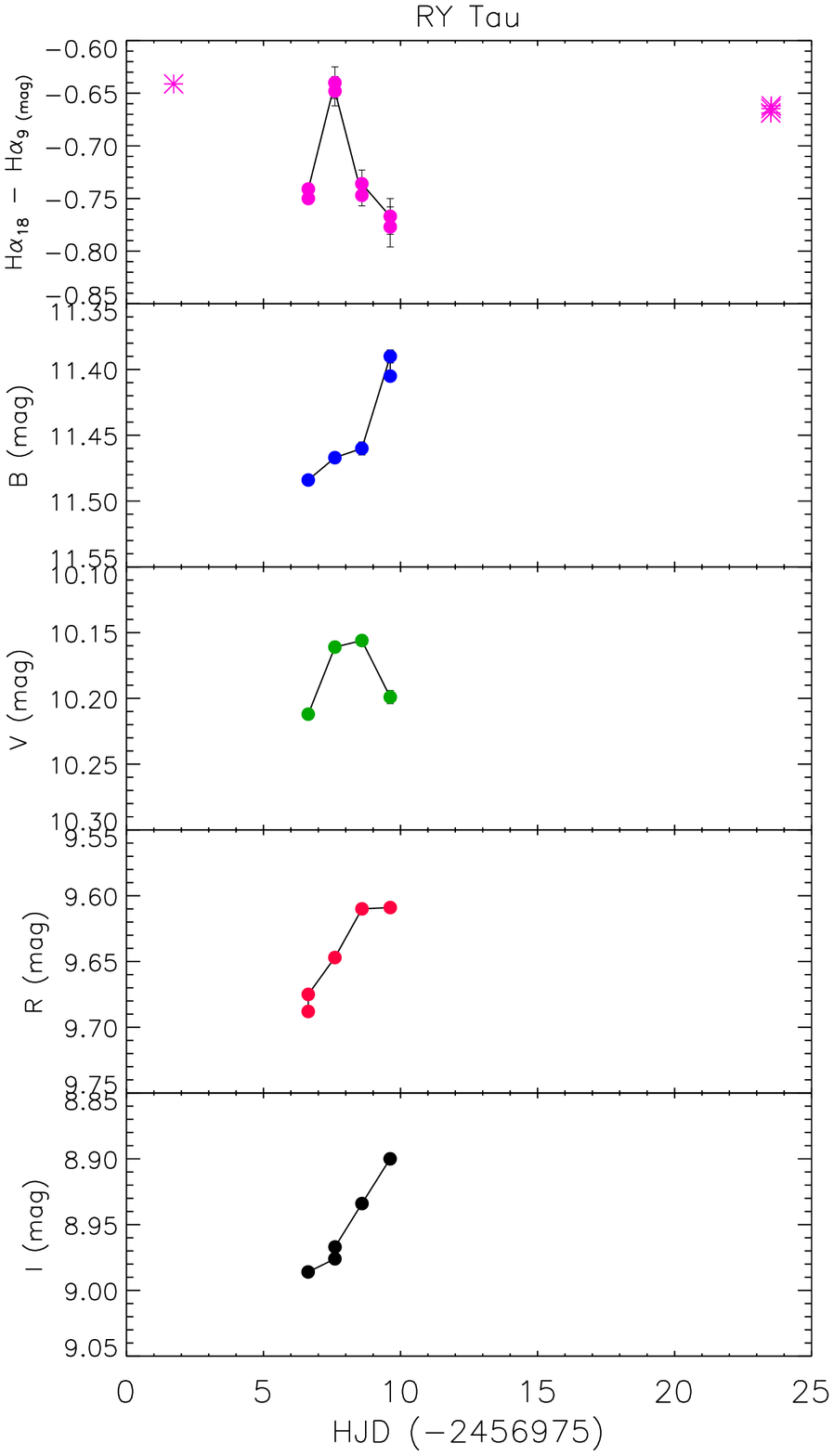}
\hspace{-0.4cm}
\includegraphics[width=4.3cm]{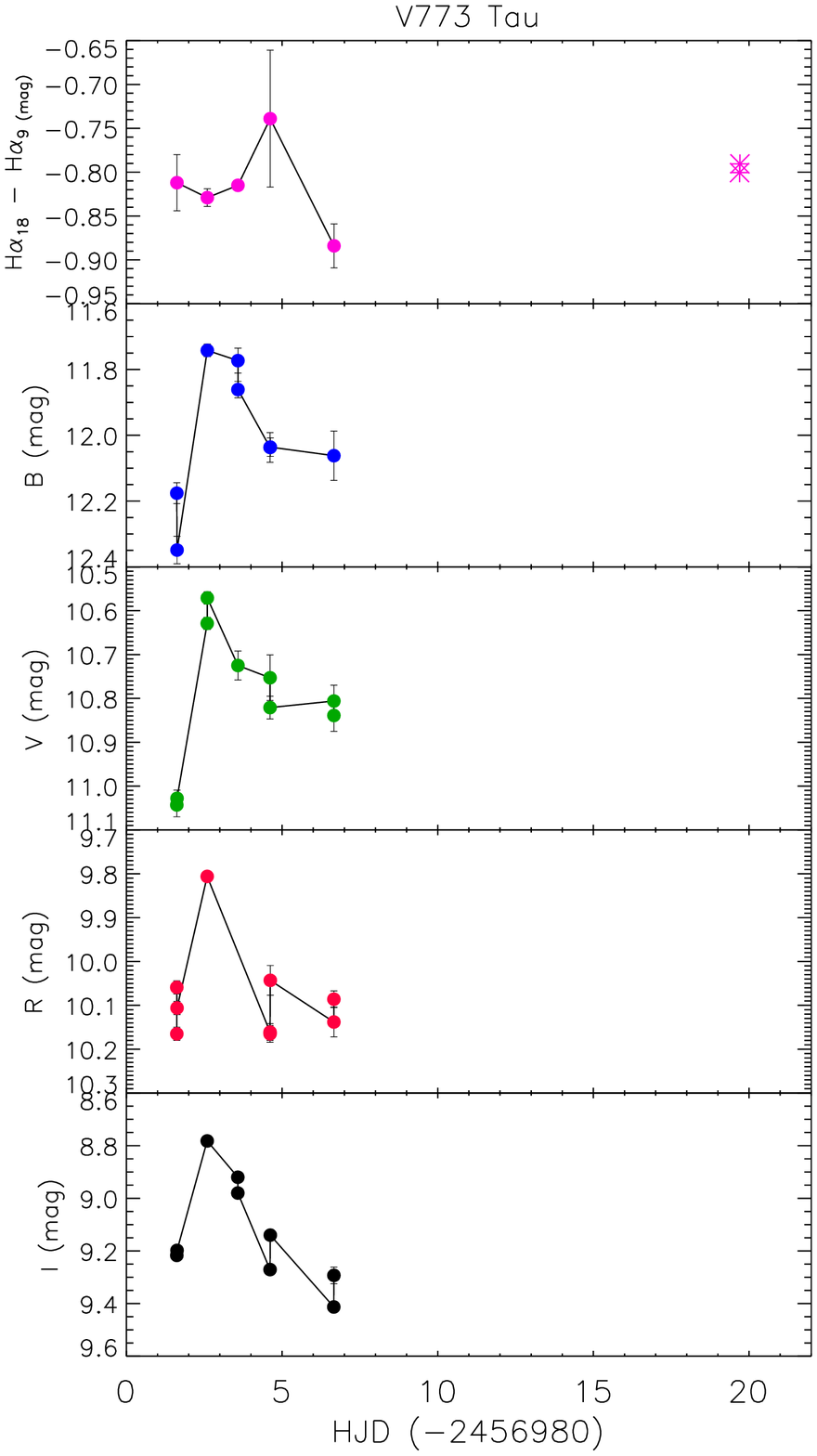}
\includegraphics[width=4.3cm]{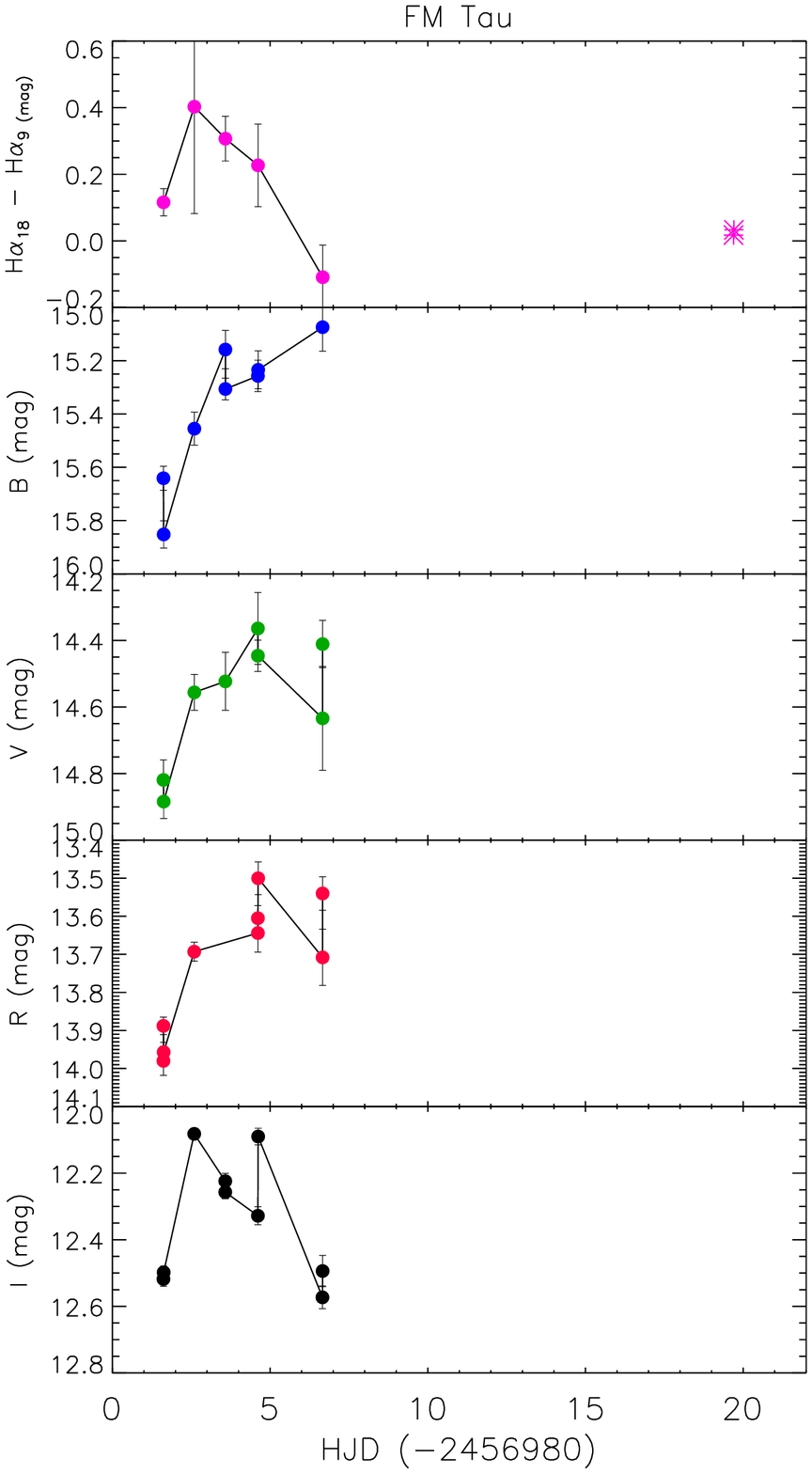}
\hspace{-0.3cm}
\includegraphics[width=4.3cm]{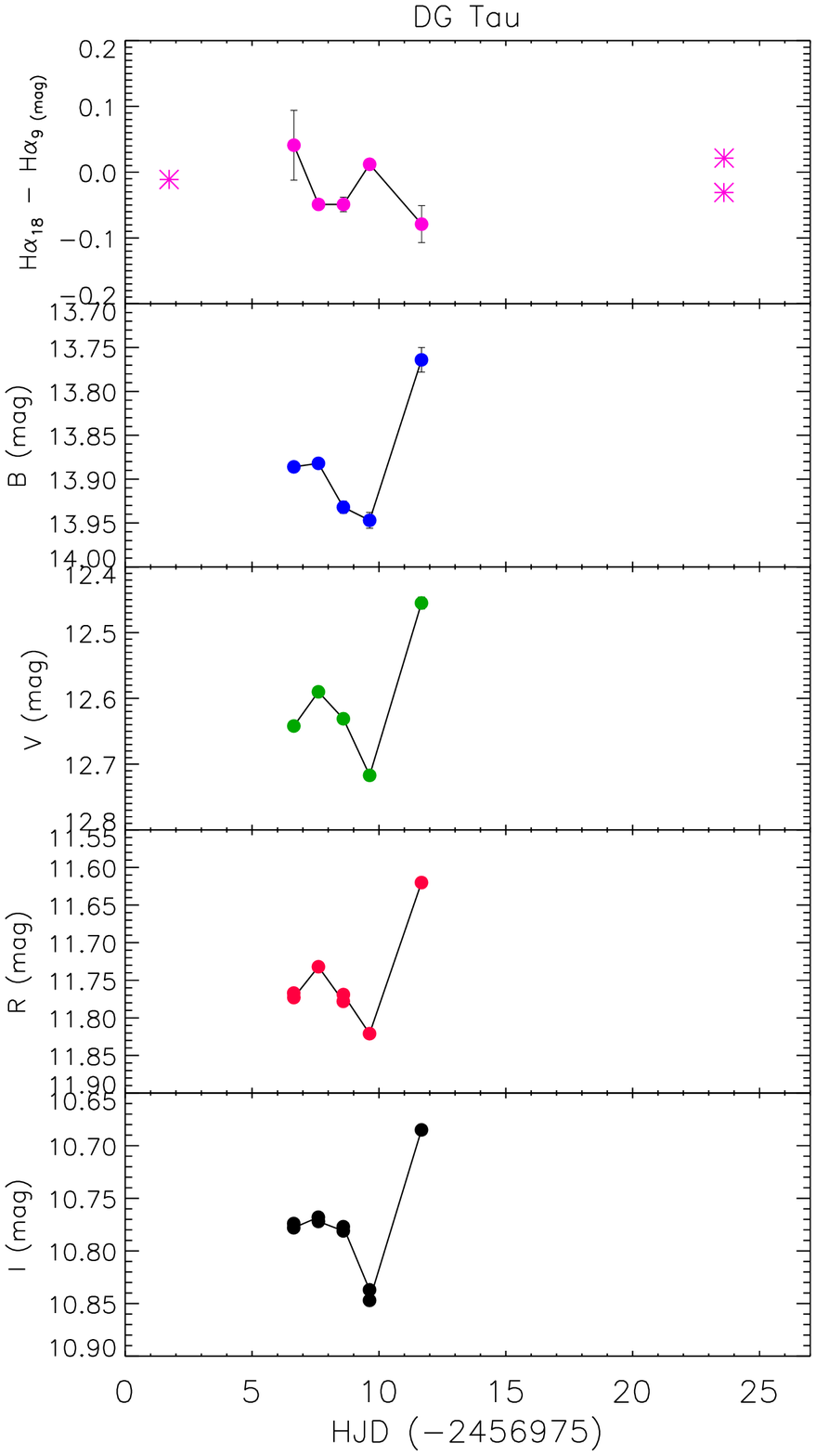}
\hspace{-0.3cm}
\includegraphics[width=4.3cm]{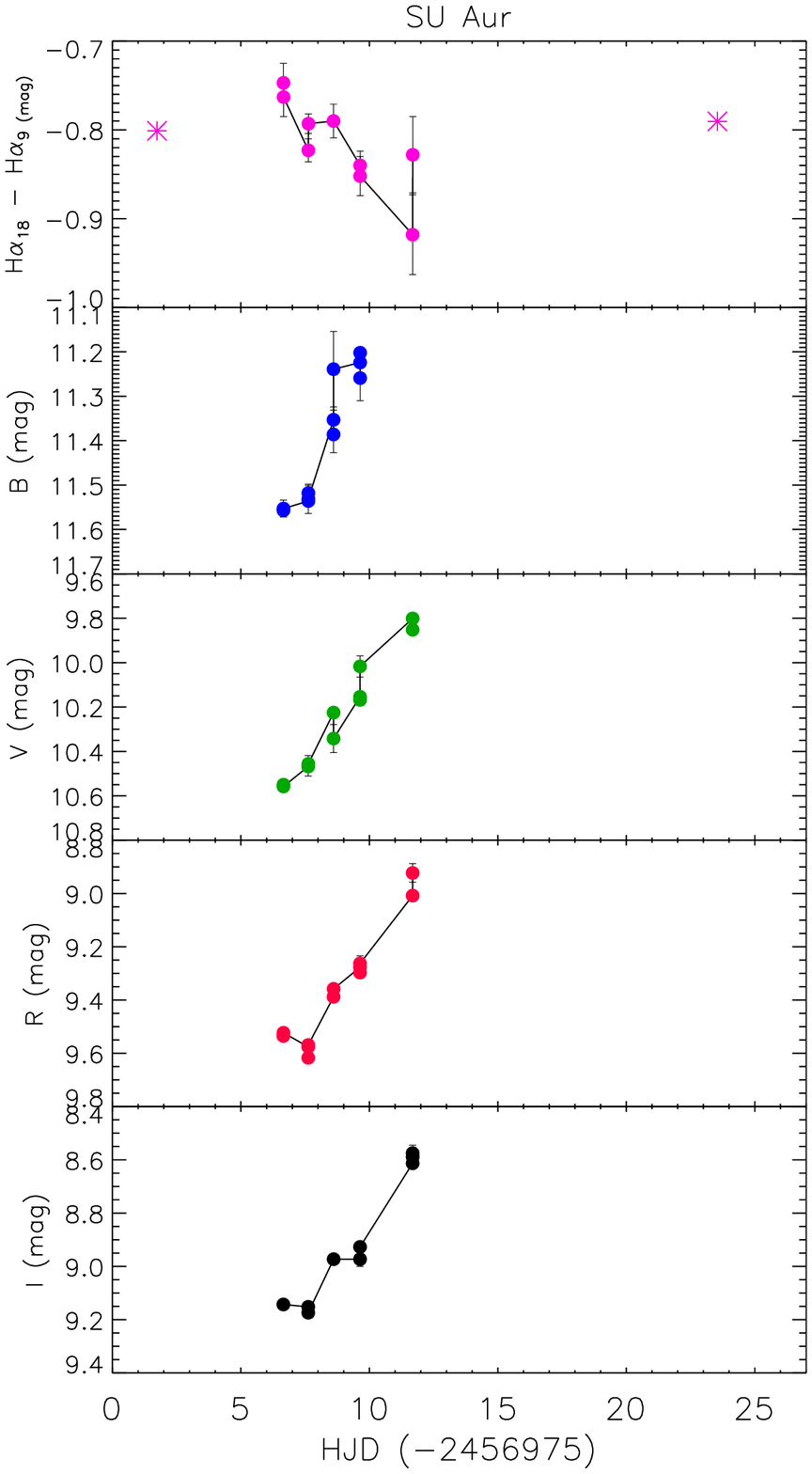}
\FigCap{{\it From left to right and from top to bottom:} Light curves of T~Tau, RY~Tau, V773~Tau, FM~Tau, DG~Tau, and SU~Aur.
In each box, from the top to the bottom panel, we plot the color index $CI_{\rm H\alpha}$, $B$, $V$, $R$, and $I$ magnitudes, respectively, as 
a function of the heliocentric Julian day. In the upper panel of each box, the dots represent the color index measured at OAC with the narrow-band filters, while the 
asterisks refer to the values measured on the IDS spectra.}
\end{figure}

Despite the short time baseline, we have observed significant variations on both broad-band and H$\alpha$ photometry for all the targets (Fig.~10).

\paragraph{\it T~Tauri --} 
The broad-band brightness shows low-amplitude (0.05--0.15\,mag) variations that could be related to the rotation of the central object ($P_{rot}\simeq 2.8$\,day)
with hot (impact shocks) or cool (magnetic activity) spots on its surface.
The H$\alpha$ index does not follow this trend, but displays a decay by about 0.08\,mag ($\approx$\,10\,\AA) from the first (JD\,=\,2\,456\,976) to the last 
(JD\,=\,2\,456\,999) day of IDS observations. 

The rotation period of about 3.0 days was also found by Percy et al. (2010) by the analysis of long-term photometric time series.
Ismailov et al. (2010) found periodicity in the equivalent widths of optical/UV emission lines, especially H$\beta$ and Ca\,{\sc ii} lines, 
on a time scale of 33\,$\pm$\,1.5 days. They note that this is not the rotation period of the star (2.8 days) and the line variation must arise 
somewhere in the magnetosphere, disk or outflow region.

\paragraph{\it RY~Tauri --}	
The brightness increases by about 0.1\,mag in $BRI$ bands while the H$\alpha$ index shows an enhancement in the second night of photometric 
observations (JD\,=\,2\,456\,982) by about 0.1\,mag ($\Delta EW_{\rm H\alpha}\approx$\,10\AA) followed by a smooth decrease.
The values of $CI_{\rm H\alpha}$ measured on the IDS spectra before and after the photometric observations are close to the maximum photometric value. 

The $V$ magnitude of about 10.2\,mag is near the median value  between 1985 and 1995 visible in Fig.~1 of Petrov et al. (1999).
These authors report also on an intensification of about one magnitude in the $V$ band in October--November 1996, ascribed to a change of opacity 
of the dusty circumstellar environment.

\paragraph{\it V773~Tauri --}
This object displays larger $BVRI$ variations than the other targets above, from 0.3\,mag up to about 0.5\,mag peak-to-peak, with a possible variation of $CI_{\rm H\alpha}$ that does not seem to be correlated with the broad-band variations.

Xiao et al. (2012) found a period of 3.075 days, based on TrEs photometry which is the same as that reported by Norton et al. (2007) and listed in 
Table~1.		
Large night-to-night X-ray variations have been found by Guenther et al. (2000) during a multi-wavelength campaign of about 10 days.
They also found correlation between the strength of H$\alpha$ and X-ray brightness.

The multiple nature of this object can be the cause of the complex behavior of variations observed in the integrated light of the 
components. A large photometric variability is reported by Boden et al. (2012), who observed variations of more than 2 magnitudes on timescales of
hundreds/thousands days in the relative A--B $K$-band photometry. They ascribe these variations to circum-B material resulting in variable line-of-sight extinction.

\paragraph{\it FM~Tauri, DG~Tauri, and SU~Aurigae --}
Large photometric variations (in the range 0.3--0.7\,mag) have been observed for the remaining three objects, namely FM~Tau, DG~Tau, and SU~Aur. 
In these cases an anticorrelation between the broad-band and H$\alpha$ photometry, i.e. the H$\alpha$ increases as the star gets fainter, 
is clearly visible. This is testified by the high values of Spearman's rank correlation coefficients $\rho$ (Press et al. 1992) between these 
quantities, which are reported in Table~4.

\setlength{\tabcolsep}{5pt}

\MakeTable{lrcrcrc}{9cm}{Rank correlation coefficients between H$\alpha$ and light curves.}
{\hline
\noalign{\smallskip}
Name       	  & \multicolumn{2}{c}{H$\alpha$--$B$} & \multicolumn{2}{c}{H$\alpha$--$V$} & \multicolumn{2}{c}{H$\alpha$--$R$} \\  
           	  &  $\rho$       &         $\sigma$   &	  $\rho$       & $\sigma$   &     $\rho$     &    $\sigma$       \\ 
\hline
\noalign{\smallskip}
T~Tau    & 0.000 & 1.000 &    0.036 & 0.938 & $-$0.109 & 0.816 \\  
RY~Tau   & 0.488 & 0.220 & $-$0.439 & 0.276 &    0.487 & 0.218 \\    
V773~Tau & 0.200 & 0.747 &    0.195 & 0.750 &    0.564 & 0.322 \\    
FM~Tau   & 0.300 & 0.623 &    0.200 & 0.747 &    0.359 & 0.553 \\
DG~Tau   & 0.616 & 0.269 &    0.872 & 0.054 &    0.875 & 0.050 \\    
SU~Aur   & 0.896 & 0.001 &    0.865 & 0.003 &    0.661 & 0.053 \\    
\hline
\noalign{\smallskip}
\multicolumn{7}{p{9cm}}{$\rho$ = Rank correlation coefficient. }\\ 	
\multicolumn{7}{p{9cm}}{$\sigma$ = Two-sided significance of the deviation from zero.}
}

Anticorrelation between the intensity of H$\alpha$ emission and light curves has been frequently observed in chromospherically active stars and 
it is ascribed to a close spatial association between dark spots in the photosphere and bright active regions in the chromosphere 
(e.g., Catalano et al. 1996, Biazzo et al. 2007, Frasca et al. 2008).
However, the four CTTs that show a high or marginal correlation between H$\alpha$ and broad-band luminosity (RY~Tau, FM~Tau, DG~Tau, and SU~Aur) 
all have shorter rotational periods than the timescales of the variations observed by us and their H$\alpha$ emission profiles are very strong and broad, 
suggesting mass accretion as their origin. 
Therefore, the observed anticorrelation is very unlikely to be the result of chromospheric/photospheric surface features, but it could rather arise from 
a variable accretion. In this hypothesis, the increase in the accretion rate indicated by the H$\alpha$ line is accompanied by a weakening of the source in 
$BVRI$, caused by the extinction produced by the material falling on the star. The opposite behavior would be observed during a decrease of the accretion
activity. 

This idea is supported by the correlations between the $V$ magnitude and colors, which are shown in Fig.~11	
for three sources.
It is interesting to note that when the star gets fainter it becomes redder in the $V$--$I$ color, but at the same time it becomes bluer in
the $B$--$V$ color. This could mean that when the accretion is stronger there is a greater extinction that shows up as a reddening of the $V$--$I$ color, 
while the $B$--$V$ color is dominated by the flux excess produced by the impact shock. 
However, this does not necessarily imply an intrinsic variation of the accretion, because the rotational modulation of accretion-related 
features (impact shock and accretion funnels) could produce similar color variations. This is probably the case for T~Tau, which displays a nice $B$--$V$ 
and $V$--$I$ anticorrelation even if the broad-band flux varies with the rotational period. 

\begin{figure}[htb]
\centering
\includegraphics[width=4.cm]{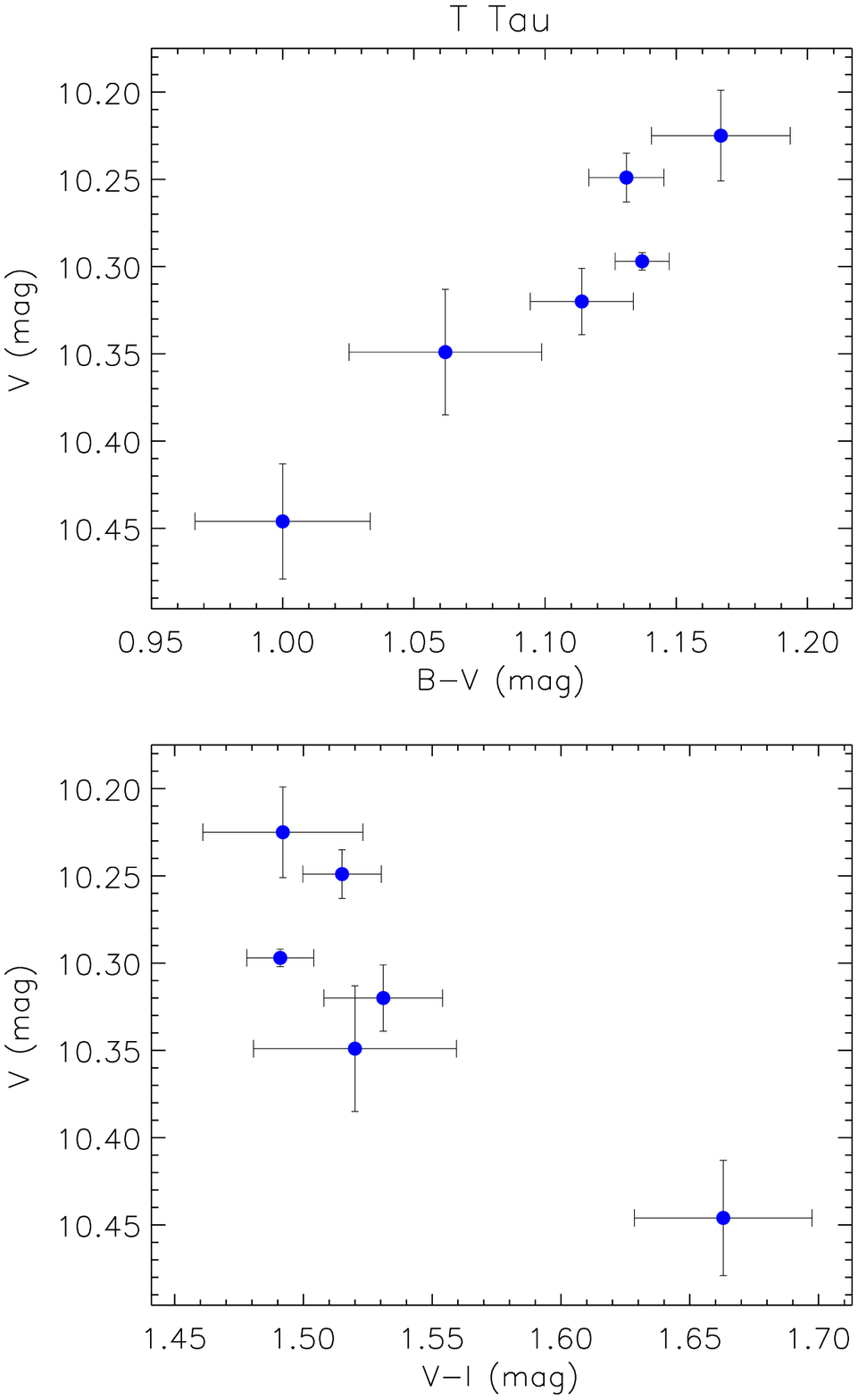}
\hspace{-0.3cm}
\includegraphics[width=4.cm]{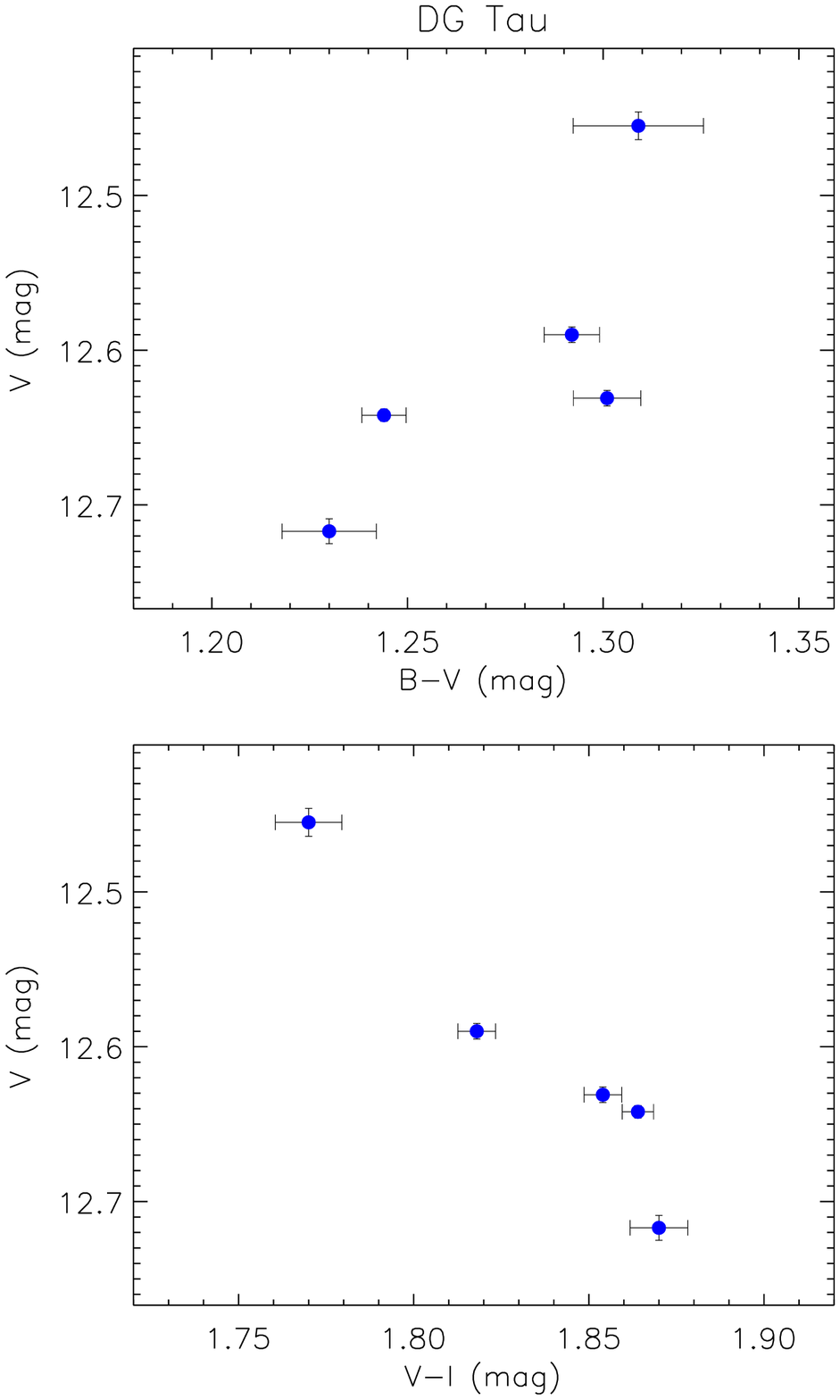}
\hspace{-0.3cm}
\includegraphics[width=4.cm]{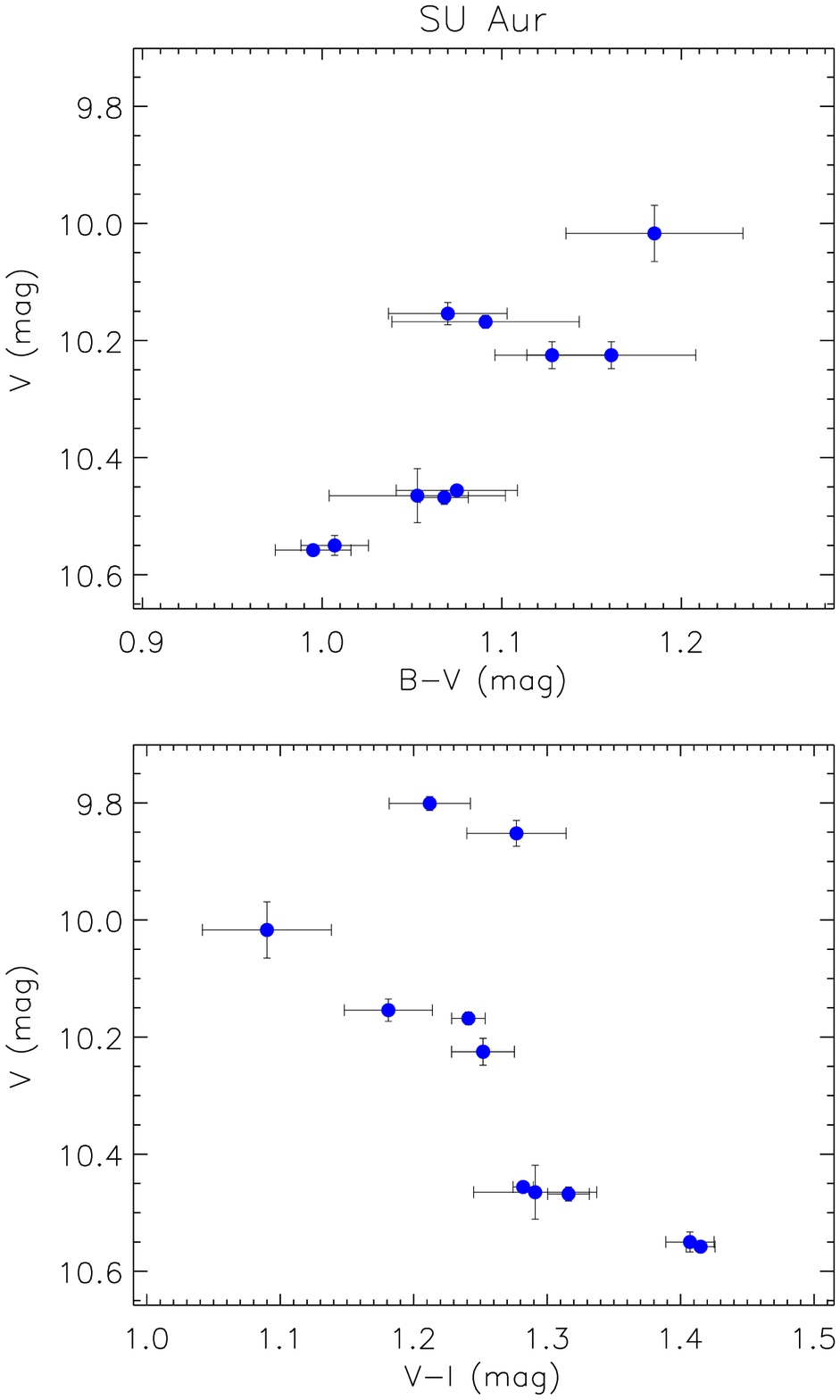}
\FigCap{$V$ magnitude as a function of the $B$--$V$ ({\it top panels}) and $V$--$I$ ({\it lower panels}) for T~Tau, DG~Tau, and SU~Aur.} 
\end{figure}

For SU~Aur the period of 2.66 days reported by Watson et al. (2015) is the same as that reported by Cody et al. (2013) on the basis of a 25-day MOST campaign.
The MOST data also displays a dimming of about 0.4\,mag during the last 5--6 days of that observing campaign.  
Our photometry shows a brightening of $\approx$\,0.6\,mag in a 6-day interval with no hint of rotational modulation. This event is accompanied by a
mild simultaneous decrease of the H$\alpha$ emission. The H$\alpha$ emission observed about twenty days later with IDS seems to be back to the initial 
high value. Unfortunately, we do not have photometric data up to this date.

\section{Discussion and conclusions}

We have presented the results of simultaneous $BVRI$  and narrow-band H$\alpha$ photometry of six T~Tauri stars in the Taurus-Aurigae complex gathered 
with the 0.9-m telescope of OAC.
Intermediate resolution spectra were also taken with IDS at the INT telescope nearly contemporaneously (before and/or after) the photometric data and 
were used both to trace the H$\alpha$ variations on a longer time scale and to derive basic stellar and accretion parameters.

Using synthetic BT-Settl and IDS spectra we carried out a characterization of the  narrow-band H$\alpha$ photometric system and obtained calibration 
relations between the equivalent width $EW_{\rm H\alpha}$ and the color index $CI_{\rm H\alpha}$, also highlighting the dependencies on other 
parameters such as \teff\ and \logg. In particular, we found that, with the typical photometric accuracy of 0.02\,mag for $CI_{\rm H\alpha}$, 
we can resolve $EW_{\rm H\alpha}$ variations at a level of 2--3\,\AA.

The analysis of the spectra allowed us to detect a significant veiling only for the sources with the largest H$\alpha$ intensities (T~Tau,
FM~Tau, and DG~Tau), as found in previous works (e.g., Fischer et al. 2011, Herczeg \& Hillenbrand 2014). 
We also inferred surface gravities in the range \logg=3.1--3.5, with the exception of the faintest target, namely FM~Tau, for which we found a rather 
uncertain value of 4.5, likely due to the low signal-to-noise spectra. These values of gravity are consistent with the very young age of 
the targets (Fig.~8).

We detected broad-band photometric variations ranging from 0.05 to 0.7 mag over the six nights of observations for all the targets.
These objects mostly display a redder $V-I$ and a bluer $B-V$ color when they become fainter. Moreover, the
sources with the largest variation amplitudes display an anticorrelation between the broad-band magnitudes and  $CI_{\rm H\alpha}$,
i.e. when the source gets fainter the H$\alpha$ intensity enhances, and vice versa. 

Despite the rather short time baseline of our photometric and spectroscopic observations ($\leq 25$ days), the timescales of broad-band 
and H$\alpha$ variations appear to be longer than 
the rotational period, with the exception of DG~Tau and T~Tau. In particular, for the latter source, the broad-band luminosity displays a clear modulation
with the rotation period, while the behaviour of the $B$--$V$ and $V$--$I$ colors seems to be related to accretion features. 
Since the strong and broad H$\alpha$ profiles of these sources are basically due to accretion, we think that chromospheric plages can be ruled out  as the
source of the observed anticorrelation between broad-band photometry and H$\alpha$ intensity. The most plausible explanation is an intrinsic variation in the 
accretion rate, at least for the objects with timescales of variations longer than the rotation period (FM~Tau and SU~Aur), while a geometric effect can
play a major role when the variation timescales are close to the rotation period. 

Dynamical instability in the magnetosphere has been invoked by Bouvier et al. (2003) to explain the photometric and spectral-line 
variations oberved on AA~Tau, while the line and continuum variations of V2129~Oph have been ascribed by Alencar et al. (2012) to rotational 
modulation of hot and cold spots plus an episodic increase of the mass accretion rate. The changes of mass accretion rates observed spectroscopically 
by Costigan et al. (2014) on a sample of Herbig Ae and T~Tau stars (including four of the sources studied in the present paper) are mainly ascribed to a
rotational modulation of the accretion-related emission, since the timescales of the variations are comparable with their rotation periods. 
Periodic H$\alpha$ variations have been observed by Sousa et al. (2016) for 8 out of 58 CTTs in the young cluster NGC\,2264, while the 
remaining do not display any clear periodicity in the line variations. Three of these sources are AA~Tau-like objects (dippers), for which the dips in the
 CoRoT light curves are ascribed to periodic occultations of the central star by a wrapped disk and the H$\alpha$ variability is likely arising from the 
geometry of the accretion columns that project over the impact region at particular rotational phases. Other sources display instead CoRoT light curves 
dominated by a spot-like rotational modulation. These cases represent a ``stable'' accretion regime
Anyway, several sources do not show any clear periodicity of the light or H$\alpha$ variations and are characterized by stochastic variations and/or bursts.
These sources are likely in a ``unstable'' regime.

Our data show variation timescales longer than the rotation period for at least two sources (FM~Tau and SU~Aur), suggesting an intrinsic 
variation of the accretion, while for the remaining sources no clear behaviour appears or the variations occur with the rotation period (as for T~Tau). 
Simultaneous photometric/H$\alpha$ observations on a longer time baseline can help to shed light on the different processes responsible for the 
observed variations and to investigate the relative importance of them.

\Acknow{We thank the anonymous referee for useful suggestions.
We acknowledge the support from {\it Regione Sicilia} and the Italian {\it Ministero dell'Istruzione, Universit\`a e  Ricerca} (MIUR).
J.M.A. acknowledges financial support from the project PRIN-INAF 2016 {\it The Cradle of Life} - GENESIS-SKA (General Conditions 
in Early Planetary Systems for the rise of life with SKA). 
D.M. acknowledges financial support from the Universidad Complutense de Madrid (UCM) and the Spanish Ministry of Economy
and Competitiveness (MINECO) from project AYA2016-79425-C3-1-P.
This research made use of SIMBAD and VIZIER databases, operated at the CDS, Strasbourg, France. 
This work has made use of data from the European Space Agency (ESA)
mission {\it Gaia} ({\tt https://www.cosmos.esa.int/gaia}), processed by
the {\it Gaia} Data Processing and Analysis Consortium (DPAC,
{\tt https://www.cosmos.esa.int/web/gaia/dpac/consortium}). Funding
for the DPAC has been provided by national institutions, in particular
the institutions participating in the {\it Gaia} Multilateral Agreement.
}


\begin{references}	
\refitem{Alcal\'a, J. M., Manara, C. F., Natta, A., et al.}{2017}{A\&A}{600}{A20}

\refitem{Allard, F., Homeier, D., \& Freytag, B.}{2012}{IAU Symp.}{282}{235}

\refitem{Alencar, S. H. P., Bouvier, J., Walter, F. M., et al.}{2012}{A\&A}{541}{A116} 

\refitem{Alfonso-Garz\'on, J., Domingo, A., Mas-Hesse, J. M., \& Gim\'enez A.}{2012}{A\&A}{548}{A79}

\refitem{Biazzo, K., Frasca, A., Henry, G. W., Catalano, S., \& Marilli, E.}{2007}{ApJ}{656}{474}

\refitem{Boden, A. F., Torres, G., Duch\^{e}ne, G., et al.}{2012}{ApJ}{747}{17}

\refitem{Bouvier, J., Cabrit, S., Fern\'andez, M., Mart\'{\i}n, E. L., \& Matthews, J. M.}{1993}{A\&A}{272}{176}

\refitem{Bouvier, J., Grankin, K. N., Alencar, S. H. P., et al.}{2003}{A\&A}{409}{169} 

\refitem{Bouvier, J., Alencar, S. H. P., Boutelier, T., et al.}{2007}{A\&A}{463}{1017}

\refitem{Brice\~no, C., Calvet, N., Kenyon, S., \& Hartmann, L.}{1999}{AJ}{118}{1354}

\refitem{Calvet, N., Muzerolle, J., Brice\~no, C., et al.}{2004}{AJ}{128}{1294}

\refitem{Catalano, S., Rodono, M., Frasca, A., Cutispoto, C.}{1996}{IAU Symp.}{176}{403}

\refitem{Cody, A. M, Tayar, J., Hillenbrand, L. A., Matthews, J. M., \& Kallinger, T.}{2013}{AJ}{145}{79}

\refitem{Cody, A. M, Stauffer, J. R., Baglin, A., et al.}{2014}{AJ}{147}{82}

\refitem{Costigan, G., Vink, J. S., Scholz, A., Ray, T., \& Testi, L.}{2014}{MNRAS}{440}{3444} 

\refitem{Davies, C. L., Gregory, S. G., \& Greaves J.S.}{2014}{MNRAS}{444}{1157}  

\refitem{Drew, J. E., Greimel, R., Irwin, M. J.}{2005}{MNRAS}{362}{753}

\refitem{Dyck, H. M., Simon, T., \& Zuckerman, B.}{1982}{ApJ}{255}{L103}

\refitem{Fischer, W., Edwards, S., Hillenbrand, L., \& Kwan, J.}{2011}{ApJ}{730}{73}

\refitem{Frasca, A., Biazzo, K., Ta{\c s}, G., Evren, S., \& Lanzafame, A. C.}{2008}{A\&A}{479}{557}

\refitem{Frasca, A., Covino, E., Spezzi, L., et al.}{2009}{A\&A}{508}{1313}

\refitem{Frasca, A., Biazzo, K., Alcal\'a, J. M., et al.}{2017}{A\&A}{602}{A33}

\refitem{Frasca, A., Biazzo, K., Lanzafame, A.~C., et al.}{2015}{A\&A}{575}{A4} 

\refitem{Furlan, E., Luhman, K. L., Espaillat, C., et al.}{2011}{ApJS}{195}{3}

\refitem{Gaia Collaboration, Brown, A. G. A., Vallenari, A., Prusti, T. et al.}{2016}{A\&A}{595}{A2} 

\refitem{Glebocki, R., \& Gnacinski, P.}{2005}{ESA, SP-}{560}{571}

\refitem{Guenther, E. W., Stelzer, B., Neuh\"auser, R., et al.}{2000}{A\&A}{357}{206}

\refitem{Gullbring, E., Hartmann, L., Brice\~no, C., \& Calvet, N.}{1998}{ApJ}{492}{323}

\refitem{Hartmann, L., Hewett, R., Stahler, S., \& Mathieu, R. D.}{1986}{ApJ}{309}{275} 

\refitem{Hartmann, L. E., Calvet, N., Gullbring, E., \& D'Alessio, P.}{1998}{ApJ}{495}{385}

\refitem{Hauck B., Nitschelm C., Mermilliod M., Mermilliod J.C.}{1990}{A\&AS}{85}{989}

\refitem{Herbig G. H., \& Bell K.R.}{1988}{Lick Observatory Bull.}{1111}{}

\refitem{Herczeg, G. J., \& Hillenbrand, L. A.}{2014}{ApJ}{786}{97}

\refitem{Huber D., Bryson S. T., Haas M. R., et al.}{2016}{ApJS}{224}{2}
	
\refitem{Ingleby, L., Calvet, N., Bergin, E., et al.}{2011}{ApJ}{743}{105}

\refitem{Ismailov, N.Z., Quliev, N.K., Khalilov, O.V., \& Herbst, W.}{2010}{A\&A}{511}{A13}

\refitem{The Hipparcos and TYCHO Catalogues}{1997}{ESA SP-}{1200}{}

\refitem{Kasper, M., Santhakumari, K. K. R., Herbst, T. M., \& K\"ohler, R.}{2016}{A\&A}{593}{A50}

\refitem{Kenyon, S. J., Hartmann, L. W., Strom, K. M., \& Strom, S. E.}{1990}{AJ}{99}{869}

\refitem{Kenyon, S. J., Dobrzycka, D. \& Hartmann, L. W.}{1994}{AJ}{108}{1872}

\refitem{Kenyon, S. J., \& Hartmann, L. W.}{1995}{ApJS}{101}{117}

\refitem{Kim, K. H., Watson, D. M., Manoj, P., et al.}{2013}{ApJ}{769}{149} 

\refitem{L\'opez-Mart\'{\i}nez, F., \& G\'omez de Castro, A. I.}{2015}{MNRAS}{448}{484}

\refitem{Manara, C. F., Testi, L., Rigliaco, E., et al.}{2013}{A\&A}{551}{A107}

\refitem{Mora A., Merin B., Solano E., et al.}{2001}{A\&A}{378}{116}

\refitem{Nguyen, D. C., Brandeker, A., van Kerkwijk, M. H., \& Jayawardhana, R.}{2012}{ApJ}{745}{119}

\refitem{Norton, A. J., Wheatley, P. J., West, R. G., et al.}{2007}{A\&A}{467}{785}

\refitem{Percy, J. R, Grynko, S., \& Seneviratne, R.}{2010}{PASP}{122}{753}

\refitem{Petrov, P. P., Zajtseva. G. V., Efimov, Yu. S., et al.}{1999}{A\&A}{341}{553}

\refitem{Press, W. H., Teukolsky, S. A., Vetterling, W. T., \& Flannery, B. P.}{1992}{Numerical Recipes in Fortran (2d ed; 
Cambridge: Cambridge Univ. Press)}{}{}

\refitem{Rebull, L. M., Padgett, D. L., McCabe, C.-E., et al.}{2010}{ApJS}{186}{259}

\refitem{Rodriguez, J. E., Ansdell, M., Oelkers, R. J., et al.}{2017}{ApJ}{848}{97}

\refitem{Siess, L., Dufour, E., \& Forestini, M.}{2000}{A\&A}{385}{593}

\refitem{Sousa, A. P., Alencar, S. H. P., Bouvier, J., et al.}{2016}{A\&A}{586}{A47} 

\refitem{Stetson, P. B.}{2000}{PASP}{112}{925}

\refitem{Valenti, J. A., Basri, G., \& Johns, C. M.}{1993}{AJ}{106}{2024}

\refitem{Venuti, L., Bouvier, J., Cody, A. M., et al.}{2017}{A\&A}{599}{A23}

\refitem{Watson, C. L., Henden, A. A., \& Price, A.}{2015}{AAVSO International Variable Star Index VSX - CDS/ADC Collection of Electronic Catalogues}{102027}{}

\refitem{Welty, A. D.}{1995}{AJ}{110}{776}

\refitem{White, R. J., \& Ghez, A. M.}{2001}{ApJ}{556}{265}

\refitem{Wright, C. O., Egan, M. P., Kraemer, K. E., \& Price, S. D.}{2003}{AJ}{125}{359}

\refitem{Xiao, H. Y., Covey, K. R., Rebull, L., et al.}{2012}{ApJS}{202}{7}

\end{references}
\end{document}